\documentclass[aip,jcp,unsortedaddress,preprint]{revtex4-1}
\usepackage{graphicx}
\usepackage{bm}
\usepackage{multirow} % required for the Table 2
\usepackage{amsmath}
\usepackage{dcolumn}% Align table columns on decimal point
\usepackage{textcomp}
\begin{document}

% Use the \preprint command to place your local institutional report
% number in the upper righthand corner of the title page in preprint mode.
% Multiple \preprint commands are allowed.
% Use the 'preprintnumbers' class option to override journal defaults
% to display numbers if necessary
%\preprint{}

%Title of paper
\title{Quasielastic small-angle neutron scattering from heavy water
solutions of cyclodextrins}

% repeat the \author .. \affiliation  etc. as needed
% \email, \thanks, \homepage, \altaffiliation all apply to the current
% author. Explanatory text should go in the []'s, actual e-mail
% address or url should go in the {}'s for \email and \homepage.
% Please use the appropriate macro foreach each type of information

% \affiliation command applies to all authors since the last
% \affiliation command. The \affiliation command should follow the
% other information
% \affiliation can be followed by \email, \homepage, \thanks as well.

\author{Andr\'e Kusmin}
\affiliation{Institute for Chemistry and Biochemistry/Crystallography, Freie Universit\"at Berlin, Taku Str. 6, 14195 Berlin, Germany }
\email[]{a.kusmin@fz-juelich.de}
%\homepage[]{Your web page}
%\thanks{}
\altaffiliation{Present address: Institut f\"ur Festk\"orperforschung, Forschungszentrum J\"ulich, 52425 J\"ulich, Germany}

\author{Ruep E. Lechner}
\affiliation{Guest at Helmholtz-Zentrum Berlin, Hahn-Meitner-Platz 1, 14109 Berlin, Germany}
% tell REL that the address changed !
\email[]{ruep.lechner@gmail.com}
%\homepage[]{Your web page}
%\thanks{}
\altaffiliation[also: ]{International Graduate College, Freie Universit\"at Berlin, Taku Str. 3, 14195 Berlin, Germany }

\author{Wolfram Saenger}
\affiliation{Institute for Chemistry and Biochemistry/Crystallography, Freie Universit\"at Berlin, Taku Str. 6, 14195 Berlin, Germany }
\email[]{saenger@chemie.fu-berlin.de}
%\homepage[]{Your web page}
%\thanks{}
%\altaffiliation{}

%Collaboration name if desired (requires use of superscriptaddress
%option in \documentclass). \noaffiliation is required (may also be
%used with the \author command).
%\collaboration can be followed by \email, \homepage, \thanks as well.
%\collaboration{}
%\noaffiliation

\date{28 October 2010}

\begin{abstract}
  % insert abstract here
We present a model for quasielastic neutron scattering (QENS) by an
aqueous solution of compact and inflexible molecules. This
model accounts for time-dependent spatial pair correlations between the
atoms of the same as well as of distinct molecules and includes all
coherent and incoherent neutron scattering contributions. The extension
of the static theory of the excluded volume effect [A. K. Soper, J. Phys.:
Condens. Matter 9, 2399 (1997)] to the time-dependent (dynamic) case
allows us to obtain simplified model expressions for QENS spectra in
the low $Q$ region in the uniform fluid approximation. The resulting
expressions describe the quasielastic small-angle neutron
scattering (QESANS) spectra of D$_2$O solutions of native and methylated
cyclodextrins well, yielding in particular translational and rotational
diffusion coefficients of these compounds in aqueous solution. Finally,
we discuss the full potential of the QESANS analysis (that is, beyond
the uniform fluid approximation), in particular, the information on
solute-solvent interactions (e.g., hydration shell properties) that such
an analysis can provide, in principle.
\end{abstract}

  % insert suggested PACS numbers in braces on next line
%\pacs{}
  % insert suggested keywords - APS authors don't need to do this
  %\keywords{}

  %\maketitle must follow title, authors, abstract, \pacs, and \keywords
\maketitle

  % body of paper here - Use proper section commands
  % References should be done using the \cite, \ref, and \label commands
\section{\label{sec:intro}Introduction}
  % Put \label in argument of \section for cross-referencing
  %\section{\label{}}
The relative significance of coherent and incoherent neutron
scattering depends on the nuclear composition of the sample, the size of the
particles or structures present in the sample, and on the range of
neutron wave vector transfer ($Q$) accessed in an experiment. In a
small-angle neutron scattering (SANS) experiment $Q$ values
are small, and, given a sufficient scattering contrast, coherent
scattering from large objects dominates the scattering pattern even
when these objects have many hydrogen nuclei (which have a high
incoherent scattering cross-section). With increasing $Q$, coherent
scattering drops fast (following for instance Guinier's law) and often becomes much
smaller than the incoherent component. Relative to SANS, in a conventional quasielastic
neutron scattering (QENS) experiment the $Q$ values are high
({\textgreater} 0.2 \AA \textsuperscript{-1}, typically {\textgreater}  0.5 \AA \textsuperscript{-1}),
the molecules are often small, and the hydrogen content is high enough to reduce the
coherent scattering contribution to a few percents and less. This is why
an analysis of QENS experiments often accounts for incoherent
scattering only (see for instance Refs.~\onlinecite{BeeQENSbook,BEE2003,SpringerLechner_DiffusionCondensedMatter}).

In a SANS diffraction experiment, incoherent scattering is just a flat background, whereas coherent
scattering is a source of structural information. In QENS, incoherent
scattering informs us about the single molecule motion, and the motions of
individual functional groups within the molecule, while coherent
scattering gives us information about the motion of molecules (and their parts)
relative to each other. 
Hence, coherent QENS is more difficult to analyze than incoherent. 
First, the dependence of the line shape of the
coherent QENS spectra on structural 
properties of the sample is more intricate; consequently, much of the structural
information (e.g., the solute's crystal structure,
radial distribution functions in solution) is required as a model
input. Second, it is in general much more difficult to model the
collective motion of a system of particles, than the motion of a single
particle.

Neutron sources and instrumentation have been and are being improved continuously, so that 
now a QENS experiment in the low $Q$ region (using longer incident
neutron wavelengths) takes a much shorter time than in the past. QENS
is increasingly often applied to study proteins and other large  
molecules. Thus, the neglect of coherent scattering in
today's QENS experiments is no longer ``automatically``
warranted. This neglect must be properly justified 
(e.g., a small contribution of coherent scattering to the total scattering cross-section does not rule out the dominance of coherent scattering in a certain $Q$ region), hence a way
to calculate (or at least estimate) the coherent scattering contribution is required. Even
more importantly, an analysis of the coherent QENS spectra may provide
unique details on the dynamics of intermolecular interactions (e.g.,
solute-water) and intramolecular interactions (e.g., internal dynamics
of proteins \cite{BUBIEHLCALLAWAY2005}).

We know few QENS studies on solutions where coherent scattering was
accounted for to some extent \cite {VASSGILANYI2005}. Neutron spin echo spectroscopy (NSE)
delivers, in principle, the same information as QENS does (although in
the time and not in the frequency domain). However, intrinsically, NSE
is more suitable for the study of coherent rather than incoherent
scattering and, relative to QENS, considerably more attention was paid
to the analysis of the former, see, e.g., 
Refs.~\onlinecite{BUBIEHLCALLAWAY2005, LONGEVILLEKALI2003}.

\begin{figure}
\includegraphics[width=8cm]{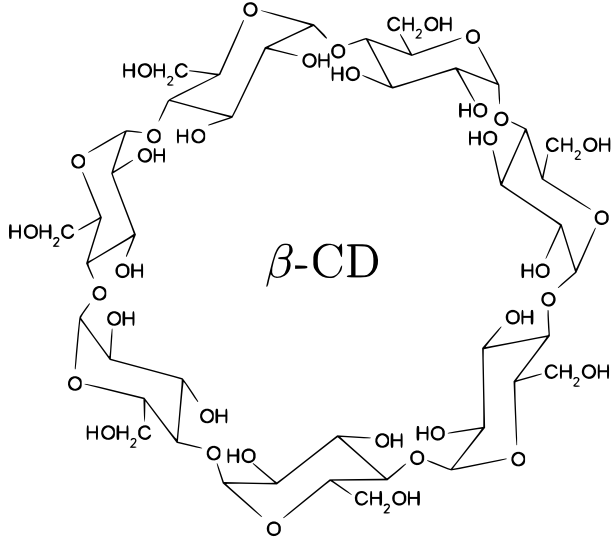}%
\caption{\label{fig:betacd}Chemical structure of $\beta$-cyclodextrin. }%
\end{figure}

Our QENS investigations \cite {KUSMINLECHNERSAENGER2007,KUSMINPHDTHESIS}  that included a partial account for
coherent scattering were on D$_2$O solutions of cyclodextrins (CDs) and
their methylated derivatives (mCDs). The CDs are macrocycles consisting
of 6, 7, or 8 D-glucose units, and are called $\alpha$-, $\beta$-,
and $\gamma$-CDs, respectively \cite{SAENGERREVIEW1998}, see Fig.~\ref{fig:betacd}. The mCDs that we studied
were $\beta$-CD per-methylated at all 2, 6 hydroxyl groups (DIMEB) and
$\gamma$-CD per-methylated at all 2, 3, 6 hydroxyl groups (TRIMEG).
While the solubility of CDs in water rises upon increasing temperature,
the opposite is true for mCDs: mCDs are well soluble in cold water but
crystallize upon heating. This makes CDs and mCDs good model systems
for the study of the hydrophobic effect and of hydration \cite{AREESAENGER2000}. In the
analysis of QENS spectra of mCD and CD solutions we calculated the
coherent scattering by a single solute molecule from atomic coordinates
known from X-ray or neutron diffraction crystal
structures, and took into account the intermolecular coherent
scattering from solutes.  Nevertheless, with these ingredients alone
we could not explain an observed excess of QENS intensity towards low $Q$
values in the spectra of mCDs dissolved in D$_2$O. We were, however,
successful in explaining this with a phenomenological model that
includes an additional coherent scattering contribution from the
hydration shell of mCDs \cite {KUSMINLECHNERSAENGER2007,KUSMINPHDTHESIS}.  In this model two approximations were made:
the coherent scattering due to solute-D$_2$O spatial correlations was
neglected, and both coherent and incoherent D$_2$O scattering in solution
were described by the same parameter values as used for the description
of the scattering by pure D$_2$O.

In this paper we develop a model for the analysis of QENS spectra of
aqueous solutions of molecules which are  relatively compact and inflexible compared to polymer
chains.  
Most significant (but not strictly necessary) assumptions of this model are: \emph{i}) hydration water is structurally and dynamically equivalent to bulk water, \emph{ii})  
scattering contributions from motions of functional groups within the solute molecule can be neglected, and \emph{iii}) the scattering function for the collective translational motion can be calculated using Vineyard's convolution approximation\cite{VINEYARD1958}. The first two assumptions   are valid for dilute solutions and $Q$ \textless 0.5 - 1 \AA\textsuperscript{-1}; the last  one is used solely for practical purposes. 
The model accounts for the time-dependent spatial 
correlations between all atoms and renders a description of all the
coherent and incoherent scattering contributions. We then extend the
static theory of the excluded volume effect\cite{SOPER1997EXCLUDEDV} to the time-dependent
case, develop simplified model expressions suitable for the QENS
spectra recorded at sufficiently low $Q$ values, and show that these
expressions are compatible with the concept of scattering contrast. 
Simplified model expressions adequately describe our QENS spectra proving that an
\emph{ad hoc} assumption about the scattering by the hydration shell made
before is not absolutely necessary (although a contribution of this
kind can not be excluded). Finally, we discuss the possibility to study
the dynamics of solute-solvent interactions by QENS.

\section{\label{sec:theory}Theory }
To help the reader in following the formulae, we have given a list of symbols at the end
of the paper, before the appendices.
  \subsection{\label{sec:theoryA}The scattering function for an aqueous solution}
  The scattering function, $S(\bm{Q},\omega)$, is the time-Fourier
  transform of the intermediate scattering function, $I(\bm{Q},t)$:
%%%
  \begin{equation}
    S(\bm{Q},\omega)=\frac{1}{2\pi }\int _{-\infty}^{\infty}e^{-i\omega
    t}I(\bm{Q},t) \, \mathrm{d} t
    \label{eq:sqwfromiqt}
  \end{equation}
%%%
  where $\bm{Q}$ is the wave vector transfer ($\bm{Q}=\bm{k}-\bm{k_0}$), 
   $\hbar\bm{Q}$ and $\hbar \omega$ are the neutron momentum and energy transfer, 
   respectively ($\hbar\omega = E - E_0$).
For an aqueous solution, $I(\bm{Q},t)$ can be written as (see, e.g., Ref.~\onlinecite{SEARS1967P3}):
%%%
\begin{equation}
  I(\bm{Q},t)=\sum _{i=1}^{N_w+N_{sol}}\sum _{\mu =1}^{}\sum _{j=1}^{N_w+N_{sol}}\sum _{\nu =1}^{}
{b_{i\mu}b_{j\nu}\left\langle e^
{-i\bm{Q}(\bm{R}_{i\mu}(0)-\bm{R}_{j\nu}(t))}\right\rangle }
  \label{eq:iqtaqueoussolution_general}
\end{equation}
%%%
where $\bm{R}_{i\mu}$ is the vector giving the position of the $\mu $th nucleus
in the $i$th molecule, $b_{i\mu}$ is the neutron scattering length of this
nucleus. $N_w$ and $N_{sol}$ denote the number of water and solute molecules,
respectively. The angle brackets denote the statistical average.

$I(\bm{Q}, t)$ can be represented as a sum of three terms depending on
solute-solute, solute-water and water-water space- and time-dependent correlations:
$I_{sol}(\bm{Q},t)$, $I_{cross}(\bm{Q},t)$, and $I_w(\bm{Q},t)$, respectively. Its time-Fourier
transform, the scattering function for an aqueous solution, is the corresponding 
sum:
%%%
\begin{equation}
  S(Q,\omega ) = S_{sol}(Q, \omega) + S_{cross}(Q, \omega ) + S_w(Q, \omega ) 
  \label{eq:sqwaqueoussolution_general}
\end{equation}
%%%
$S_{sol}(Q, \omega )$ accounts for the intermolecular coherent scattering
(due to the time-dependent pair-correlations between the positions and
orientations of two distinct solute molecules) and for the intramolecular
scattering (due to self-correlations between the positions and orientations the single solute molecule takes on at different times). The latter generally is
a sum of coherent and incoherent scattering. Likewise, $S_w(Q, \omega )$
accounts for the intermolecular coherent scattering, and the
intramolecular (coherent and incoherent) scattering from water
molecules. Finally, $S_{cross}(Q, \omega )$ accounts for the
intermolecular coherent scattering due to solute-water time-dependent
spatial correlations (there is no incoherent scattering contribution
here because the correlations are between different molecules).

For the solute molecules that are relatively compact and inflexible (as
opposed to linear polymers, alkanes etc) and just as well for bulk water molecules, the model given in
Appendix \ref{sec:appendixA} can be used to calculate the scattering functions, 
i.e.,  $S_{sol}(Q, \omega )$ and $S_{w}(Q, \omega )$, respectively. 
This model was
originally developed for molecular liquids, and the motion of an atom located in
a molecule is described by the convolution of the center-of-mass (CM)
diffusion of the molecule and an isotropic rotational diffusion. The shape of the
 molecule does not have to be spherical, it should just not be too
anisotropic. For flexible molecules other models should be used (e.g.,
for polymers: CM diffusion and the Rouse model).
In the following we write down, as an example, the detailed expression for $S_{w}(Q, \omega )$; the formally completely analogous expression for $S_{sol}(Q, \omega )$ can then be obtained simply when replacing everywhere the subscript \emph{w} by the subscript \emph{sol}.

For the moment, we assume  that hydration water and bulk water are
structurally and dynamically equivalent. Consequently, the expressions from Appendix \ref{sec:appendixA} 
(Eqs.~(\ref{eq:appA:total})-(\ref{eq:appA:totallall})) 
can be used for $ S_w(Q, \omega )$. Explicitly, $ S_w(Q, \omega )$
is:
%%%
\begin{subequations}
  \begin{equation}
    S_{w}(Q,\omega )=n_{w}\text{DWF}_{w}\sum _{l=0}^{\infty }S_{w}^{l}(Q,\omega )    
    \label{eq:sqw_water_1}
  \end{equation}
  \begin{equation}
    S_{w}^{0}(Q,\omega)=A_{w}^{0~coh}(Q) S^{coh}_{tr~w}(Q,\omega)+A_{w}^{0~inc}(Q)S_{tr~w}^{inc}(Q,\omega ) \quad
 l = 0
    \label{eq:sqw_water_2}
  \end{equation}
  \begin{equation}
   S_{w}^{l}(Q,\omega )=(2l+1)\left ( A_{w}^{l~coh}(Q)+A_{w}^{l~inc}(Q) \right )S_{tr~w}^{inc}(Q,\omega)
   {\otimes} \text{Lor} (l(l+1)D_{r~w},\omega ) \quad l \neq 0  
    \label{eq:sqw_water_3}
  \end{equation}
\end{subequations}
%%%
where $n_w$ is the number density of water molecules in solution, DWF$_w$ is
the Debye-Waller factor, ${\otimes}$ is the convolution operator, and
$D_{r~w}$ is the rotational diffusion coefficient of a water molecule. The
coefficients $A^l_w(Q)$ are given by Eq.~(\ref{eq:appA:AlQcoef}), Lor $(x, \omega )$ stands
for a Lorentzian function with $\omega $ and $x$ being the argument and
the parameter (half-width at half maximum), respectively. $S^{coh}_{tr~w}(Q, \omega )$ and
$S^{inc}_{tr~w}(Q, \omega )$ are the coherent and incoherent translational
scattering functions for the CM of water molecules (corresponding to 
 collective diffusion and self-diffusion of water molecules, respectively).

The expression for $S_{cross}(Q, \omega )$ can be written as (see Appendix 
\ref{sec:appendixB}):
%%%
\begin{equation}
   S_{cross}(Q,\omega)=2n_{sol}b_{sol}(Q)b_{w}(Q)S_{tr~sol-w}(Q,\omega )
  \label{eq:sqwcrossterm}
\end{equation}
%%%
where $n_{sol}$ is the solute number density, $b_{sol}(Q)$ and $b_w(Q)$ are the
effective scattering lengths (see Eq.~(\ref{eq:appA:beffQdefine})) of solute and water
molecules, respectively. $S_{tr~sol-w}(Q, \omega )$ is the (coherent)
scattering function for the translational motion of water molecules
relative to solute molecules.

We made one standard, difficult to avoid, assumption: the
rotational and translational motions of a molecule, as well as
rotational motions of two distinct molecules, are not coupled (weak
hindering approximation \cite{SEARS1967P3}). Throughout the paper we will also assume
that, because of the low-$Q$ region of our experiment,
 rotational motions of water molecules and  motions of functional groups within the solute molecule contribute to the QENS spectra to a negligible extent. Although this is not strictly true, we make this assumption 
because our primary goal is to consider the intermolecular coherent scattering, which is only observable in the low-$Q$ region
($Q$ {\textless} 0.5 - 1 \AA \textsuperscript{-1}).

There exist a number of  theoretical  expressions for the incoherent translational
scattering function, but none for the coherent one. That is, there are
no expressions for $S^{coh}_{tr~sol}(Q, \omega )$, $S^{coh}_{tr~w}(Q, \omega )$ and $S_{tr~sol-w}(Q, \omega )$,
but we need them to use Eq.~(\ref{eq:sqwaqueoussolution_general}). 
For our present practical purpose, since we do not have a fully valid theory at our disposal, 
we proceed by applying Vineyard's convolution approximation \cite{VINEYARD1958}. 
Although this approximation has no profound theoretical justification, it is a means of constructing 
an at least phenomenologically approximate coherent scattering function $S^{coh}_{app}(Q, \omega )$ from an incoherent translational scattering function $S^{inc}(Q, \omega )$ by multiplying the latter with the known $Q$-dependent integral $S(Q)$ of $S^{coh}(Q, \omega )$. 
By doing this, the 0\textsuperscript{th} moment of $S^{coh}_{app}(Q, \omega )$  becomes correct, which does however not imply the correctness of the higher moments of $S^{coh}_{app}(Q, \omega )$; (see also Refs.~\onlinecite{TassoSpringerQENS1972CONVAPP,LechnerMassTransportSolids1983CONVAPP} for some more information about this). In Vineyard's  approximation we have:
%Refs.~\cite[(\S10.2)][]{TassoSpringerQENS1972}, \cite[(pp. 219-220 in)][]{LechnerMassTransportSolids1983} 
%\onlinecite[§10.2]{TassoSpringerQENS1972:\S10.2}
%\textsuperscript{,}
%%%
\begin{equation}
  S^{coh}_{tr~sol}(Q,\omega )=S_{cm~sol}(Q)S_{tr~sol}^{inc}(Q,\omega)
  \label{eq:convapp_solute}
\end{equation}
\begin{equation}
  S^{coh}_{tr~w}(Q,\omega )=S_{cm~w}(Q)S_{tr~w}^{inc}(Q,\omega)
  \label{eq:convapp_water}
\end{equation}
\begin{equation}
    S_{tr~sol-w}(Q,\omega ) \approx S_{sol-w}(Q)S_{tr~w}^{inc}(Q,\omega) 
  \label{eq:convapp_solw}
\end{equation}
%%%
where $S_{cm~sol}(Q)$, $S_{cm~w}(Q)$, and $S_{sol-w}(Q)$ are the solute-solute,
water-water, and solute-water intermolecular CM structure factors in
solution (hereafter: structure factors). The solute-solute structure
factor can be obtained from a small-angle neutron, X-ray or light
diffraction experiment or can be calculated (see, e.g., Refs.~\onlinecite{KINNINGTHOMAS1984,HAYTER1981}). The
solute-water and water-water structure factors can, in principle, be
obtained from neutron diffraction experiments \cite{SOPERTURNER1993INTJMODPHYS}. The solute-water
structure factor can also be calculated; a way to do this is shown in
Appendix \ref{sec:appendixC}.  Note an approximate equality in  Eq.~(\ref{eq:convapp_solw});  
 this is explained in Appendix \ref{sec:appendixC}, too.

Above we assumed that hydration water and bulk water are
structurally and dynamically equivalent. If the structure of hydration
water differs from the bulk, three additional structure factors are
required for the correlations: [hydration water -- bulk water],
[hydration water -- solute], and [hydration water -- hydration water].
Fortunately, at $Q$ values where intermolecular coherent scattering is
important, a (slight) difference in the hydration water structure is
likely to have no influence on the scattering from solution. Only a
(slightly) different number density of water molecules in the hydration
shell may have to be taken into account (via $n_w$ in Eq.~(\ref{eq:sqw_water_1})). On the
other hand, the dynamics of hydration water may be substantially
different from that of bulk water. Still, $S_{w}(Q, \omega )$ from Eqs.~(\ref{eq:sqw_water_1})-(\ref{eq:sqw_water_3}), and
$S_{cross}(Q, \omega )$ from Eq.~(\ref{eq:sqwcrossterm}) will remain applicable, if we use 
 the convolution approximation and take for $S^{inc}_{tr~w}(Q, \omega)$ a two
state model, e.g., the model of Singwi and Sj\o
lander \cite{SINGWISJOELANDER1960}. In a two
state model, both translational and rotational dynamics could be
described by two sets of parameters, for the bulk and for the hydration water, respectively.
Note that although the rotational dynamics of the hydration water is different
 from that of the bulk water, in our $Q$ region the effect of using somewhat different
rotational diffusion coefficients is small (the terms for $l > 0$ are negligible).

Thus, the framework described above makes it possible to account for the coherent scattering.
\subsection{\label{sec:theoryB} Uniform fluid approximation}
In general, for a practical application of the approach described above all three
structure factors from Eqs.~(\ref{eq:convapp_solute})-(\ref{eq:convapp_solw}) have to be known. 
However, when $Q$ values are sufficiently  small, one can use the approximation of the solvent by a uniform continuum (hereafter: the
uniform fluid approximation or the UFA);
in this case,  
 $S_{cm~w}(Q)$ and $S_{sol-w}(Q)$ are not required. 
 Below we derive the corresponding $S^{coh}_{tr~w}(Q, \omega )$- and $S_{tr~sol-w}(Q, \omega )$-expressions and show that 
 they depend only on the solute structure and dynamics.

The intermediate scattering function, $I(\bm{Q},t)$, is the
space-Fourier transform of the time-dependent pair correlation
function, $G(\bm{r},t)$:
%%%
\begin{equation}
I(\bm{Q},t)=\int _{V}e^{i\bm{Qr}}G(\bm{r},t)\, \mathrm{d}\bm{r}
  \label{eq:iqtfromgrt_general}
\end{equation}
%Thus, $S_{TR~W}(Q, \omega )$ and $S_{TR~SOL-W}(Q, \omega )$ can be obtained from the corresponding pair correlation functions.
Using the uniform fluid approximation A. K. Soper derived the expressions for the
static pair correlation functions in solutions, $G_{HH}(r)$, $G_{XX}(r)$ and
$G_{XH}(r)$, where $H$ is  a hydrogen atom in a solvent molecule       and $X$ is any atom in a solute molecule  \cite{SOPER1997EXCLUDEDV}\textsuperscript{,}
\footnote{
We note that, using Soper's notations and definitions, we arrived at expressions for $g_u^{(XX)}(r)$ and $g_u^{(XH)}(r)$ that differed slightly from those given by Eq. (20) in the original paper. Specifically, $g_u^{(XX)}(r)$ in our result has the factors $(1+f_S)^2 V_p/V$ and ${\{(1+f_S)V_p/V\}}^2$ instead of $f^2_S V_p/V $ and $ { \{f_S V_p/V\} }^2$, respectively, whereas $g_u^{(XH)}(r)$ has the factors $ -(1+f_S)V_p/V $ and $-(1+f_S)V_p^2/V^2 $ instead of $-f_S V_p/V $ and $ -f_S V_p^2/V^2$, respectively.
}.
For our purposes we derived similar expressions by taking the CM of a water molecule as $H$ and the CM of a solute molecule as $X$. We further extended  Soper's
approach to obtain the expressions for  the time-dependent translational water-water and
 solute-water pair correlation functions, $G_{tr~w}(r,t)$ and $G_{tr~sol-w}(r,t)$ 
 (hereafter: $G_{w}(r,t)$ and $G_{sol-w}(r,t)$), respectively
(see Appendix \ref{subsec:generalformalism}-\ref{subsec:uniformapplication}).
The expressions for $S^{coh}_{tr~w}(Q, \omega )$ and $S_{tr~sol-w}(Q, \omega )$ 
follow from the time-Fourier transformation of 
%The space-Fourier transforms of $G_{w}(r,t)$ and $G_{sol-w}(r,t)$
the intermediate scattering functions ($I^{coh}_{tr~w}(Q,t)$ and $I_{tr~sol-w}(Q,t)$, respectively) 
that are given in Appendix \ref{subsec:uniformiqt}.

In the UFA $G_{w}(r,t)$ reflects the
time-dependent spatial pair correlations between two
(infinitesimal) volume elements of the solvent, and $G_{sol-w}(r,t)$
reflects such correlations between the CM of the solute molecule and
the solvent volume element. These correlations depend on the
translational motion of the solute molecules relative to each other,
described by $G_{sol}(r, t)$, and, if the solute molecules do not have a
spherical shape, on their rotational motion described by $G_{sol}^{dist
(p)}(r, t)$ and $G_{sol}^{self(p)}(r, t)$. The superscript $(p)$ indicates 
the function's relation to the volume element inside a
particle (in our case, inside a solute molecule). Specifically,
$G_{sol}^{dist(p)}(r, t)$ describes the orientational  correlation of the volume
element of the solute molecule with another volume element of a
distinct solute molecule at a different time; $G_{sol}^{self (p)}(r, t)$
describes the orientational correlations between the volume elements of
the same solute molecule.

$S^{coh}_{tr~w}(Q, \omega )$ is the time-Fourier transform of $I^{coh}_{tr~w}(Q,t)$  defined by  Eqs.~(\ref{eq:appD:itrw}), (\ref{eq:appD:xirotation}):
%%%
\begin{subequations}
  \begin{equation}
    S^{coh}_{tr~w}(Q,\omega )=\frac{n_{sol}}{n_{w}}\sum
_{l=0}^{\infty }S^{l(p)}(Q,\omega )
    \label{eq:sqwwater_uniform1}
  \end{equation}
  \begin{equation}
    S^{0(p)}(Q,\omega)=A^{0(p)}(Q)S^{coh}_{tr~sol}(Q,\omega ) \quad l = 0
    \label{eq:sqwwater_uniform2}
  \end{equation}
  \begin{equation}
S^{l(p)}(Q,\omega)=(2l+1)A^{l(p)}(Q)S_{tr~sol}^{inc}(Q,\omega)
{\otimes}\text{Lor}(l(l+1)D_{r~sol},\omega ) \quad l \neq 0
    \label{eq:sqwwater_uniform3}
  \end{equation}
\end{subequations}
where the coefficients $A^{l(p)}(Q)$ are given by Eq.~(\ref{eq:appD:alpcoefficients}). 
%%%

$S_{tr~sol-w}(Q, \omega )$ is  the time-Fourier transform of $I_{tr~sol-w}(Q,t)$ given by Eq.~(\ref{eq:appD:itrsolw}):
% the solute-water translational scattering function,
%
\begin{equation}
  S_{tr~sol-w}(Q,\omega)=-N^{(p)}(Q)S^{coh}_{tr~sol}(Q,\omega)
  \label{eq:sqwcross_uniform}
\end{equation}
where $N^{(p)}(Q)$ is given by Eq.~(\ref{eq:appD:NPQ}).

In Eqs.~(\ref{eq:sqwwater_uniform1}-\ref{eq:sqwwater_uniform3}) and (\ref{eq:sqwcross_uniform}) 
the quasielastic broadening depends on the solute structure and dynamics only.
Furthermore, the $S^{coh}_{tr~w}(Q, \omega )$-expression 
is similar to that for $S_{sol}(Q, \omega )$ (see Appendix \ref{sec:appendixA}). This is a consequence of the uniform
fluid approximation: the solvent has no structure, and therefore solvent volume elements effectively do not move themselves.

Now that $S^{coh}_{tr~w}(Q, \omega )$ and $S_{tr~sol-w}(Q,\omega )$ are
derived, the scattering function for an aqueous solution is fully
defined by Eqs.~(\ref{eq:sqwaqueoussolution_general}) - (\ref{eq:sqwcrossterm}). For a practical application, we still need some means to calculate
 the coherent translational solute
scattering function in Eqs.~(\ref{eq:sqwwater_uniform2}) and (\ref{eq:sqwcross_uniform}); here we use Eq.~(\ref{eq:convapp_solute}).

While the uniform fluid approximation neither affects the
calculation of incoherent scattering, nor that of the 
coherent scattering for $l = 1, 2 \cdots$ in Eq.~(\ref{eq:sqw_water_3}), it underestimates the term for 
coherent scattering for $l = 0$ in Eq.~(\ref{eq:sqw_water_2}). This term accounts for the intermolecular coherent scattering of water molecules in solution and reads:
\begin{equation}
  S_{w}^{inter}(Q,\omega )=A_{w}^{0~coh}(Q)S^{coh}_{tr~w}(Q,\omega)
  \label{eq:sqwintermoleculard2ogeneral}
\end{equation}
%%
%$S_{w}^{inter}(Q,\omega )$ 
In the UFA, $S^{coh}_{tr~w}(Q, \omega )$ is given by Eq.~(\ref{eq:sqwwater_uniform1}); it does not depend on the local water structure and water dynamics, but it \emph{does} depend on the change in the water structure caused by the volume excluded by solute molecules.
%, and therefore it contains, implicitly, the intermolecular water structure, $S_{cm~w}(Q)$.
%To see, approximately, what kind of scattering contribution is neglected, 
Without the UFA,   Eq.~(\ref{eq:sqwintermoleculard2ogeneral}) can be rewritten (using Vineyard's convolution approximation) as:
\begin{equation}
S_{w}^{inter}(Q,\omega )= S_{cm~w}(Q)A_{w}^{0~coh}(Q)S_{tr~w}^{inc}(Q,\omega) 
  \label{eq:sqwintermoleculard2o}
\end{equation}
In Eq.~(\ref{eq:sqwintermoleculard2o})  $S_{cm~w}(Q)$ depends on both the local water structure and the presence of solute molecules; the line broadening of $S_{w}^{inter}(Q,\omega )$ depends on water dynamics. Thus, we see that the UFA 
does not account for the broad coherent scattering component due to translational water dynamics. The intensity 
of this component can be estimated from the coherent scattering of pure water 
in the low-$Q$ region; 
as known from experiment, 
 in many cases it is negligible compared to all other scattering contributions,
especially for non-dilute solutions. 
\subsection{\label{sec:theoryC}Low $Q$ limit and scattering contrast}
Even in the uniform fluid approximation the expression for the total
scattering function for an aqueous solution is quite involved. Let us
find a simplified expression in the limit of very low $Q$ values and
without the incoherent scattering contribution.

Qualitatively, one expects rotational and vibrational motions to
have little effect in QENS spectra at low $Q$ values. Specifically, at $Q$
{\textless} 0.5 \AA \textsuperscript{-1}, the $A_{w}^{l}(Q)$-coefficients for $l \neq 0 $ are
negligible and DWF$_{w}{\approx}$ 1, so Eq.~(\ref{eq:sqw_water_1}) can be written as:
\begin{equation}
  S_{w}(Q,\omega )=n_{w}b_{w}^{2}(Q)S^{coh}_{tr~w}(Q,\omega)
  \label{eq:lowqwatercomponent}
\end{equation}
where $b_{w}(Q)$ is defined by Eqs.~(\ref{eq:appA:beffQdefine})-(\ref{eq:appA:beffqa0coh}).
Likewise, at a sufficiently low $Q$ value (which depends on the size of
the solute molecule) $A^{l}_{sol}(Q)$ and $A^{l(p)}(Q)$ for $l \neq  0 $ are
negligible, too. Therefore, Eqs.~(\ref{eq:appA:totall0}), (\ref{eq:appA:beffqa0coh}), and (\ref{eq:sqwwater_uniform1}) yield:
\begin{equation}
  S_{sol}(Q,\omega)=n_{sol}b_{sol}^{2}(Q)S^{coh}_{tr~sol}(Q,\omega )
  \label{eq:lowqsolcomponent}
\end{equation}
\begin{equation}
  S^{coh}_{tr~w}(Q,\omega)=\frac{n_{sol}}{n_{w}}A^{0(p)}(Q)S^{coh}_{tr~sol}(Q,\omega )
  \label{eq:lowqtranswater}
\end{equation}
Thus, the coherent QENS scattering from solution is: 
\begin{equation}
  S^{coh}(Q,\omega)=n_{sol}\left (  b_{sol}^{2}(Q) -2b_{sol}(Q)b_{w}(Q)N^{(p)}(Q) + A^{0(p)}(Q)b_{w}^{2}(Q) \right ) S^{coh}_{tr~sol}(Q,\omega ) 
  \label{eq:lowqsqwtotal}
\end{equation}
At low $Q$, $N^{(p)}(Q)$ is just the number of water
molecules excluded from the solution by one solute molecule, and
$A_{0}^{(p)}(Q)$ is equal to the square of this number, as follows from 
Eqs.~(\ref{eq:appD:alpcoefficients}), (\ref{eq:appD:NPQ}). Thus, Eq.~(\ref{eq:lowqsqwtotal}) yields  (after applying the convolution
approximation):
\begin{equation}
S^{coh}(Q,\omega)=n_{sol}\left (b_{sol}(Q)-b_{w}(Q)N^{(p)}(Q)\right )^{2} S_{cm~sol}(Q)S_{tr~sol}^{inc}(Q,\omega)
  \label{eq_lowqsqwtotal2}
\end{equation}

While Eq.~(\ref{eq_lowqsqwtotal2}) is already simple enough for a practical application,
it can be simplified further to make its meaning more transparent.
Let $\rho _{sol}$ and $\rho _w$ be the solute and water coherent scattering
length densities (in general, $\rho _{mol} = b_{mol}(Q\to 0)/V_{mol}$). Then $\rho _{sol} - 
\rho _{w}$ is the scattering contrast and  Eq.~(\ref{eq_lowqsqwtotal2}) can be written as:
\begin{equation}
 S^{coh}(Q,\omega )=n_{sol}(\nu ^{(p)})^{2}(\rho _{sol}-\rho _{w})^{2}S_{cm~sol}(Q)S_{tr~sol}^{inc}(Q,\omega)
  \label{eq:lowqsqwtotal3}
\end{equation}
where $\nu ^{(p)}$ is the volume excluded by the solute molecule. As follows from
Eq.~(\ref{eq:sqwfromiqt}), an integration of Eq.~(\ref{eq:lowqsqwtotal3}) over energy transfer gives the
small-angle scattering intensity, $I(Q,t=0)$. Because $\int S_{tr~sol}^{inc}
(Q, \omega )\, \mathrm{d} \omega {\equiv} 1 $, we get an equation that is well-known
in the field of small-angle neutron and X-ray scattering:
\begin{equation}
  I^{coh}(Q,t=0)=n_{sol}(\nu ^{(p)})^{2}(\rho _{sol}-\rho _{w})^{2}S_{cm~sol}(Q)
  \label{eq:iqtcontrastconverge}
\end{equation}
This result demonstrates that the model framework presented in this paper is, generally speaking,
 an account for the scattering contrast in the
time-dependent case. 
\section{\label{sec:experiment}Experiment}
\subsection{\label{sec:experimentA}Experimental Details}
D$_2$O 99.9 \% pure, DIMEB and TRIMEG ({\textgreater} 95\%, CycloLab),
$\gamma $-CD ({\textgreater}98\%, ROTH) were used without further
purification. In our calculations the density of the solutions was
taken to be equal to the density of pure D$_2$O because the solute
concentrations were low.  %

QENS spectra of D$_2$O and of solutions of DIMEB (50 mg/mL), TRIMEG
(61.4 mg/mL) and $\gamma $-CD (48.7 mg/mL) in D$_2$O, were recorded with
the time-of-flight (TOF) spectrometer NEAT at BENSC,
Hahn-Meitner-Institut (HMI, Berlin) by one of us (REL). The sample
containers were circular slabs with thicknesses of 1.6 or 2.5 mm,
the sample transmission was {\textgreater} 0.85 (with the beam
perpendicular to the slab). %

In one experiment the spectra were recorded with an energy
resolution ($\Delta E$), full width at half maximum, of ${\approx}$ 10
$\mu $eV. The incident neutron wavelength ($\lambda _0$) was 10.0 \AA,
sample angle ($\alpha $)\footnote{
The sample angle $\alpha$ is the angle between the incident beam and the plane of the sample slab,  0\textdegree $\leq \alpha \leq $ 180\textdegree. For  $\alpha$=45\textdegree, the 90\textdegree-scattering is in reflection geometry, for  $\alpha$=135\textdegree, the 90\textdegree-scattering is in transmission geometry. 
For  $\alpha$=0\textdegree, the sample plane would be parallel to the incident beam. 
%(hence this angle is not used in practice).
%
   }=90\textdegree, the range of the $Q$ values for zero
energy transfer ($Q$ range, hereafter) was from 0.16 \AA \textsuperscript{-1} to 1.2 \AA \textsuperscript{-1}.
In another experiment, the spectra were recorded with $\Delta E
{\approx}$ 10 $\mu $eV, $\lambda _0$=15.3 \AA, $\alpha$=60\textdegree, the
$Q${} range was from 0.10 \AA \textsuperscript{-1} to 0.75 \AA \textsuperscript{-1}. For the QENS analysis the
$Q$ range was limited by a maximum value of ${\approx}$0.6 \AA \textsuperscript{-1}, in order to remain in the
low $Q$ region. The sample temperatures are given in Tables \ref{tab:tabled2o} and \ref{tab:tablecds}.

%TABLES
%TABLES
%TABLES
\begin{table}
\caption{\label{tab:tabled2o} Parameter values for the D$_2$O scattering model: 
$D_{tr~w}$ = translational diffusion coefficient of water molecules; 
$\tau _{tr~w}$ = translational diffusion correlation time in this model; 
$\left <  u^2 \right > _{w}$ = mean-square displacement; 
$D_{r~w}$  = rotational diffusion coefficient of water molecules;  
these parameter values were taken from the literature, 
see Section \ref{sec:experimentB} for details.}
\begin{tabular}{lccccc}
  \hline
  \hline
  ~ &
T [K] & 
$D_{tr~w}$ [$10^{-5}$ cm\textsuperscript{2}/s] &
$\tau _{tr~w}$ [ps] &
$\left <  u^2 \right > _{w}$ [{\AA}\textsuperscript{2}] &
 $D_{r~w}$  [$\mu$eV]\\
  \hline

$\lambda _{0}=10 $ \AA   &
288 &
1.389 &
0.75 &
0.077 &
88.96\\
  
\cline{1-6}
\multirow{3}{*}{$\lambda _{0}=15.3 $ \AA} &
285.5 &
1.294 &
0.827 &
0.077 &
86.47\\
 ~&
301 &
1.970 &
0.502 &
0.077 &
102.3\\
  
 ~ &
317.7 &
2.958 &
0.400 &
0.077 &
120.4\\
  \hline
  \hline
\end{tabular}
\end{table}
\begin{table}
\caption{\label{tab:tablecds}  Solute translational ($ D_{tr~sol}$ [$10^{-5}$ cm\textsuperscript{2}/s]) and
rotational ($ D_{r~sol}$ [$\mu$eV]) diffusion coefficients in D$_2$O solutions. The
values with uncertainties were obtained by fitting the model to the QENS spectra.}
\begin{tabular}{llccccc}

  \hline
  \hline
  ~ &
  Sample &
T [K] & 
$D_{tr~sol}$ \footnote{From the literature, see Section \ref{sec:experimentB}.} &
$D_{tr~sol}$ \footnote{Fitted with $D_{r~sol}$ fixed at 0. }&
$D_{tr~sol}$ \footnote{Fitted with $D_{r~sol}$ fixed at 0.25 $\mu$eV} &
$D_{r~sol}$  \footnote{Fitted with $D_{tr~sol}$ fixed at  the values from column 4.}\\
\hline
\multirow{3}{*}{ $\lambda _{0}=10 $ \AA } &
$\gamma $-CD &
303.6 &
0.268 &
0.504{$\pm$}0.016 &
0.393{$\pm$}0.016 &
0.55{$\pm$}0.03 \\
 ~ & 
TRIMEG &
308 &
0.280 &
0.458{$\pm$}0.012 &
0.383{$\pm$}0.011 &
0.56{$\pm$}0.03\\
 ~ & 
DIMEB &
303 &
0.184 &
0.240{$\pm$}0.008 &
0.180{$\pm$}0.009 &
0.32{$\pm$}0.03\\

\hline
\multirow{9}{*}{$\lambda _{0}=15.3 $ \AA } &
\multirow{3}{*}{$\gamma $-CD} &
285.5 &
0.144 &
0.367{$\pm$}0.015 &
0.263{$\pm$}0.014 &
0.59{$\pm$}0.04\\

 ~ & 
 ~ & 
301 &
0.246 &
0.465{$\pm$}0.016 &
0.362{$\pm$}0.016 &
0.59{$\pm$}0.05\\

 ~ & 
 ~ & 
317.7 &
0.416 &
0.506{$\pm$}0.021 &
0.408{$\pm$}0.021 &
0.28{$\pm$}0.06\\

\cline{2-7}

 ~ & 
 \multirow{2}{*}{TRIMEG} &
285.5 &
0.118 &
0.140{$\pm$}0.005 &
0.087{$\pm$}0.004 &
0.13{$\pm$}0.02 \\
   &
 ~ & 

300.8 &
0.216 &
0.204{$\pm$}0.004 &
0.150{$\pm$}0.004 &
0.009{$\pm$}0.02\\

\cline{2-7}
 ~ & 
 \multirow{4}{*}{DIMEB} &
278.1 &
0.083 &
0.106{$\pm$}0.005 &
0.062{$\pm$}0.005 &
0.25{$\pm$}0.03\\
&
 &
290.8 &
0.126 &
0.120{$\pm$}0.006 &
0.081{$\pm$}0.006 &
0.17{$\pm$}0.03\\

 &
 &
303.7 &
0.186 &
0.148{$\pm$}0.006 &
0.108{$\pm$}0.006 &
0.05{$\pm$}0.04\\
 &
 &
317 &
0.268 &
0.287{$\pm$}0.008 &
0.226{$\pm$}0.008 &
0.29{$\pm$}0.05\\
\hline
\hline
\end{tabular}
\end{table}
\subsection{\label{sec:experimentB}Data Analysis}
Data reduction of the raw QENS spectra was carried out using the program FITMO \footnote{FITMO is a program package for QENS data reduction and analysis. It can be requested from M. Russina, Helmholtz Centre Berlin (former Hahn-Meitner-Institut). }. The energy resolution function was
determined by fitting a Gaussian function to the vanadium spectra. The
expression fitted to the sample spectra reads:  
\begin{equation}
S(Q, \omega) = F_{sc}(\phi )e^{-\hbar \omega /2 k_B T}S(Q, \omega )\otimes R(\phi , \omega )
  \label{eq:sqfit}
\end{equation}
where $\phi $ is the scattering angle, $R(\phi , \omega )$ --- slightly angle-dependent 
energy resolution function, $e^{-\hbar \omega /2 k_B T}$ ---
detailed balance factor, $k_B$ --- Boltzmann constant, $T$ --- absolute
temperature, $F_{sc}(\phi )$ --- scaling factor, $S(Q, \omega )$ ---
theoretical scattering function in the classical approximation.

The $S(Q, \omega )$-expression fitted to the QENS spectra of pure
D$_2$O, as well as the $S_{sol}(Q, \omega )$-expression describing the
scattering by solute molecules are defined in Appendix \ref{sec:appendixA}. In the
spectra analyzed here, $Q$ {\textless} 0.6 \AA \textsuperscript{-1}; for this low-$Q$ range, \emph{i}) the DWF of the solute can be approximated by
unity (for water, taking $\left < u^2 \right >_w$ from Ref.~\onlinecite{TEIXEIRADIANOUX1985}, the
DWF decays to 0.97 at $Q$=0.6 \AA \textsuperscript{-1}, and it is reasonable to expect a significantly
smaller value of $\left < u^2 \right >_{sol}$), \emph{ii})  
 the influence of the $\tau _{tr~sol}$-value on the translational
diffusion line width given by Eq.~(\ref{eq:appAfqtrans}) is negligible. Consequently, for
the contribution of translational diffusion to $S_{sol}(Q, \omega )$ we used (instead
of Eqs.~(\ref{eq:appA:sqwtrinc}),(\ref{eq:appAfqtrans})):
\begin{equation}
  S^{inc}_{tr~sol}(Q, \omega ) = \text{Lor}(D_{tr~sol}Q^2, \omega )
  \label{eq:specialsqwsolute}
\end{equation}
which is the well-known form of expression (\ref{eq:appA:sqwtrinc}) in the low-$Q$ limit.

The  literature sources for pure-D$_2$O parameter values were: Refs. \onlinecite{MILLS1973,WILBURJONAS1976,DEFRIES1977} for $D_{tr~w}$, Ref. \onlinecite{KUSMINPHDTHESIS} for $\tau _{tr~w}$;
$D_{r~w}$ and $\left < u^2 \right >_w$ originate from studies on H$_2$O\textsuperscript{~} \cite{TEIXEIRADIANOUX1985}. See Table
\ref{tab:tabled2o} for the values actually used in fits to the QENS spectra of pure D$_2$O
and D$_2$O solutions. $S_{cm~w}(Q)$ was calculated in the ''static
approximation'' (see, e.g., Eq.~(14) in Ref.~\onlinecite{SEARS1967P3}) from the D$_2$O data (O-D
bond length, D-O-D angle and the intermolecular function $D_M(Q)$) taken
from neutron diffraction \cite{MCBFTEIXEIRA1991}. Because in our solutions the solute volume
fraction was less than 0.05, we neglected the change in the D$_2$O
diffusion coefficient compared to that of pure D$_2$O. From the crystal
structures of $\gamma $-CD, DIMEB, and TRIMEG \cite{AREEHOIER1999,AREESAENGER2000P2,DINGSAENGER1991} we computed the
$A^l_{sol}(Q)$-coefficients and, using the cube method\cite{MUELLER1983}, the functions
$N^{(p)}(Q)$ and $A^{l(p)}(Q)$. Van der Waals (vdW) radii were taken as 1.75,
1.58, and 1.1 \AA{} for C, O, and H atoms, respectively \cite{ROWLANDTAYLOR1996}. To account for
the difference between the molecule's vdW volume
and the volume excluded by the molecule, a shell of thickness $\Delta $
around the vdW volume was used ($\Delta $=0.1, 0.26, and 0.33 \AA{}  for $\gamma $-CD, DIMEB and TRIMEG, respectively\cite{KUSMINSAENGER2008}). More details on
our implementation of the cube method are given elsewhere \cite{KUSMINSAENGER2008}. A multiple
scattering calculation was carried out  at every iteration of the non-linear
least squares fitting procedure as previously described \cite{KUSMINPHDTHESIS}.

The $D_{tr~sol}$ values of $\gamma $-CD, DIMEB, and TRIMEG\cite{UEDAIRAUEDAIRA1990,GAITANOCD1997}\textsuperscript{,}\footnote{PFG-NMR measurements by D. Leitner, private communication.  } were
corrected for the differences in viscosity of H$_2$O relative to D$_2$O \cite{CHOROBINSON1999}.
The $D_{tr~sol}$ data used here (Table \ref{tab:tablecds}, column 4) were found by
inter- and extrapolation of the literature values  using the Arrhenius law for the temperature dependence and an analogous exponential law for the concentration dependence. 
From NMR results, for $\beta $-CD in D$_2$O  at 25\textcelsius, 
the rotational correlation time  $\tau _{r~sol}$ is 220 ps \cite{SALVADORICD1993},
 corresponding to $D_{r~sol}$=0.5 $\mu$eV  (according to $D_{r~sol}$[meV]=0.6583/6$\tau _{r~sol}$[ps]).
 Since molecules we studied are larger than $\beta $-CD, smaller $D_{r~sol}$ are expected. 
Therefore, the $D_{r~sol}$ values, if not
fitted, were kept at 0.25 $\mu$eV, 0.1 $\mu$eV or 0 $\mu$eV; the quality of the fits differed negligibly.

%(approximately corresponding
%to $\tau _{R~SOL}$ = 220 ps as reported \cite{SALVADORICD1993} for $\beta $-CD, 2 mM in D$_2$O at
%25$^\circ$C)

To calculate $S_{cm~sol}(Q)$, we extended Debye's
approach for the calculation of $S_{cm~sol}(Q)$ for hard spheres to the case
of hard bodies of an arbitrary shape. We assumed that given a molecule
with an orientation $\Omega _1$ and its center-of-mass (CM) at the
origin, the probability to find the CM of another molecule with an
orientation $\Omega _2$ at a distance r is equal to the mean solute
number density everywhere, as long as molecular volumes do not overlap,
and zero otherwise. The static CM pair correlation function $G_{cm~sol}(r,
\Omega _1, \Omega _2)$ was calculated using the cube method, averaged
over all possible orientations $\Omega _1$ and $\Omega _2$, and Fourier
transformed to yield $S_{cm~sol}(Q)$. Although the so-obtained $S_{cm~sol}(Q)$ 
accounts for the two-body interactions only, it is adequate given the
low solute volume fraction in the studied solutions (see, e.g., Ref.~\onlinecite{KINNINGTHOMAS1984}).

If the correction procedure for the angle-dependent attenuation of
the incident beam and of the sample scattering is accurate, and the
spectra were normalized to the scattering by vanadium, the scaling factor $F_{sc}(\phi )$ is just a
constant that can be calculated from the sample thickness and the
properties of the calibration standard 
%\cite{RieutordINXManual} 
\footnote{F. Rieutord, ``INX -- Program for time-of-flight data reduction'' (ILL, 1990). It can be requested from the Time-of-flight and High-Resolution Group at ILL or downloaded from ILL's internet site.}. 
However, due to the
approximations used in the correction procedure, $F_{sc}(\phi )$ 
usually deviates from the expected value. To compensate for those
small, but non-negligible deviations in the fitting procedure of the
model expressions to the spectra, $F_{sc}(\phi )$ was employed as a
free but $\phi $-dependent fitting parameter.

Since the UFA does not account for the intermolecular coherent scattering due to a finite size
of water molecules, $S_{w}^{inter}(Q,\omega )$ from Eq.~(\ref{eq:sqwintermoleculard2ogeneral}) lacks a broad scattering component which we call $S_{w}^{corr}(Q,\omega ) $. We estimate the magnitude of this component by the intermolecular coherent scattering from pure water, which is
 (see Eq.(\ref{eq:sqwintermoleculard2o})):
\begin{equation}
S_{w}^{corr}(Q,\omega )= S_{cm~d2o}(Q)A_{w}^{0~coh}(Q)S_{tr~w}^{inc}(Q,\omega) 
  \label{eq:sqwcorr}
\end{equation}
where  $S_{cm~d2o}(Q)$ is the structure factor for pure D$_2$O. 
At low $Q$, 
where only the first term in the infinite series (\ref{eq:sqw_water_1}) needs to be considered, 
the ratio of   the energy-integrated function  $S_{w}^{corr}(Q,\omega )$  to  the energy-integrated incoherent water scattering  
(see Eq.~(\ref{eq:sqw_water_2})) is: $ S_{cm~w}(Q)A^{0~coh}_{w}(Q)/A^{0~inc}_{w}(Q)$=0.79 ~
  \footnote{
To estimate $S_{cm~w}(Q)A^{0~coh}_{w}(Q)/A^{0~inc}_{w}(Q)$ we used:
$A^{0~coh}_{d2o}$($Q$=0.6  \AA $^{-1}) \approx $ 3.7 barn (cf. Eq.~(\ref{eq:appA:AlQcoef})), $S_{cm~d2o}$($Q$\textless0.6  \AA $^{-1})$ is below 0.07~ \cite{MCBFTEIXEIRA1991}, $A^{0~inc}_w(Q) \approx 2 \sigma _D/4\pi \approx 0.33 $ barn. Please note that, unlike $A^{0~coh}_{d2o}(Q)$, the structure factor $S_{cm~d2o}(Q)$ is dimensionless. }. 
Therefore, we decided that for the dilute
solutions (as in our case) it was better to add $S_{w}^{corr}(Q,\omega )$ to the model expression given by the UFA  than to neglect it entirely. Hence, in fitting of the
QENS spectra a modified version of Eq.~(\ref{eq:sqw_water_2}) was used:
\begin{equation}
  S_{w}^{0}(Q,\omega)=A_{w}^{0~coh}(Q)S^{coh}_{tr~w}(Q,\omega)+  
A_{w}^{0~inc}(Q) S_{tr~w}^{inc}(Q,\omega)  + S_{w}^{corr}(Q,\omega )
  \label{eq:sqwinterapproximated}
\end{equation}
\section{\label{sec:results}Results}
  %
  %	end of experimental section
  %
  %
  %
  % to the QENS spectra of CD and mCD solutions  
The model fitted to all QENS spectra of CD and mCD solutions is represented by 
Eq.~(\ref{eq:sqwaqueoussolution_general})  containing  the sum of the three terms 
$S_{w}(Q,\omega )$, 
$S_{cross}(Q,\omega )$ and $S_{sol}(Q,\omega )$. The function $S_{w}(Q,\omega )$ is given by 
Eqs.~(\ref{eq:sqw_water_1})-(\ref{eq:sqw_water_3}), where Eq.~(\ref{eq:sqwinterapproximated}) is replacing (\ref{eq:sqw_water_2}),
and by (\ref{eq:sqwwater_uniform1})-(\ref{eq:sqwwater_uniform3}); $S_{cross}(Q,\omega )$ is given by Eqs.~(\ref{eq:sqwcrossterm})  
and (\ref{eq:sqwcross_uniform}), 
while $S_{sol}(Q,\omega )$ is given by Eqs.~(\ref{eq:sqw_water_1})-(\ref{eq:sqw_water_3}) 
(with superscript ``sol'' instead of ``w``), (\ref{eq:convapp_solute}) 
and (\ref{eq:specialsqwsolute}).

\begin{figure}[htp]
\centering
\includegraphics[width=0.45\textwidth]{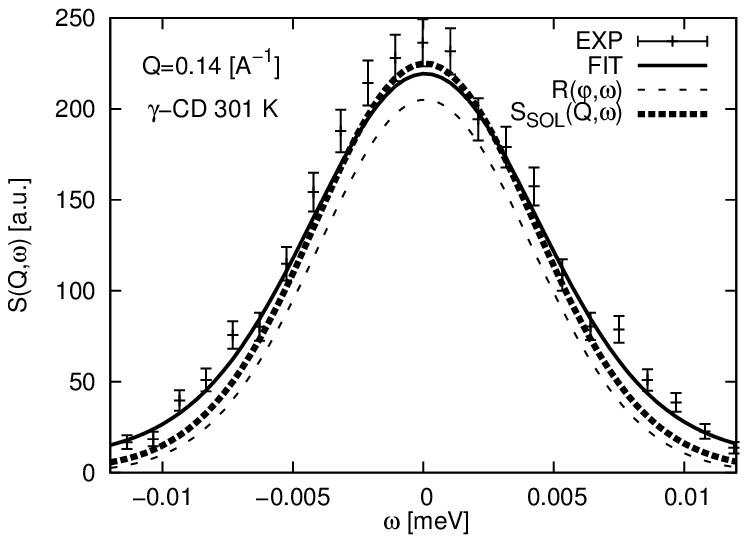}
\hfill
\includegraphics[width=0.45\textwidth]{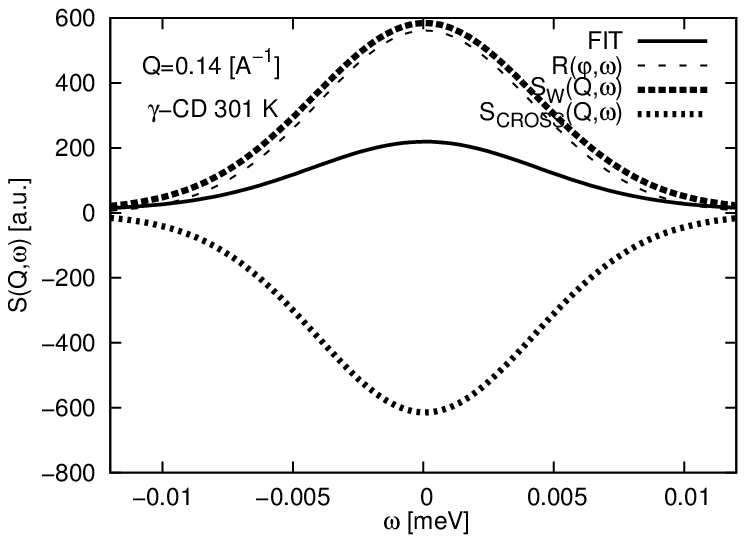}\\
\includegraphics[width=0.45\textwidth]{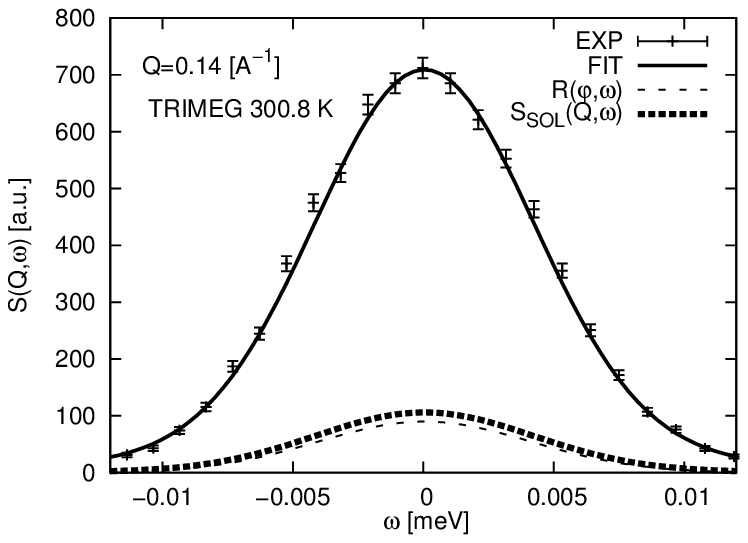}
\hfill
\includegraphics[width=0.45\textwidth]{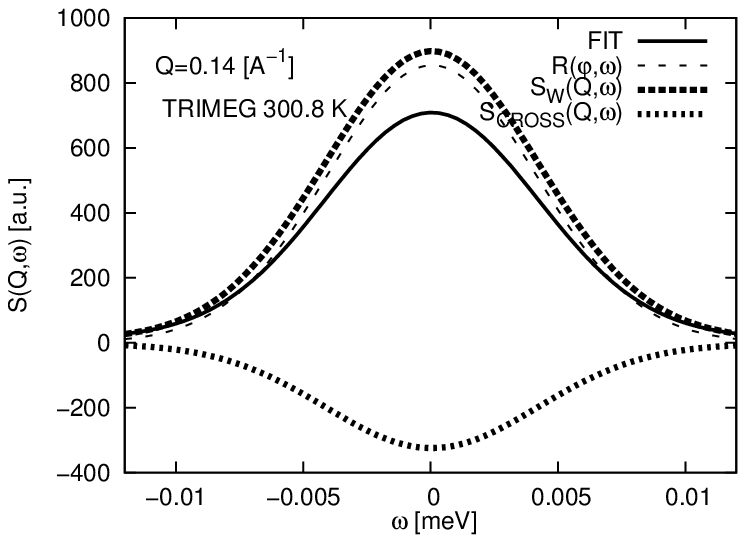}\\
\includegraphics[width=0.45\textwidth]{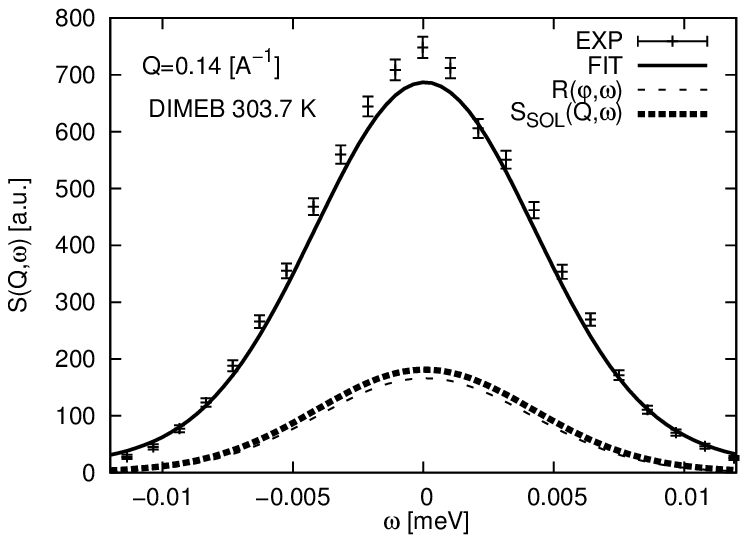}
\hfill
\includegraphics[width=0.45\textwidth]{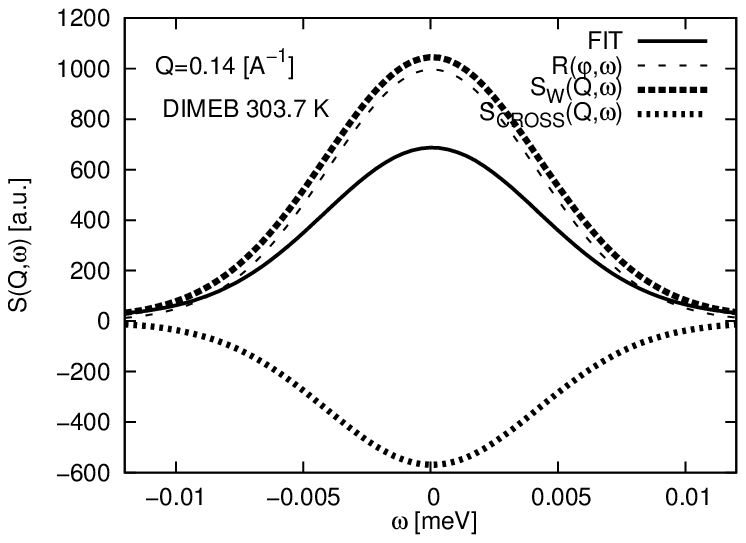}\\
\caption{Examples of fitting of the model to the QENS spectra of cyclodextrins dissolved in D$_2$O, for the experiment with $\lambda _0$=15.3 \AA. ``EXP'' and ``FIT'' stand for experimental data and the fitted curve, respectively; the ``FIT''-curve is the same in both columns. $S_{sol}(Q, \omega)$ and $S_w(Q, \omega)$ are the solute and D$_2$O scattering, respectively (both coherent plus incoherent); $S_{cross}(Q, \omega)$ is the coherent scattering due to D$_2$O-solute time-dependent spatial correlations. The energy resolution function, $R(\phi, \omega)$, is plotted for the comparison of instrumental broadening with the broadening of the separate scattering contributions. 
For the theoretical origin of the scattering functions  $S_w(Q, \omega)$,  $S_{cross}(Q, \omega)$, and  $S_{sol}(Q, \omega)$, see the beginning of this section.}\label{fig:fitlowq}
\end{figure}

\begin{figure}[htp]
\centering
\includegraphics[width=0.45\textwidth]{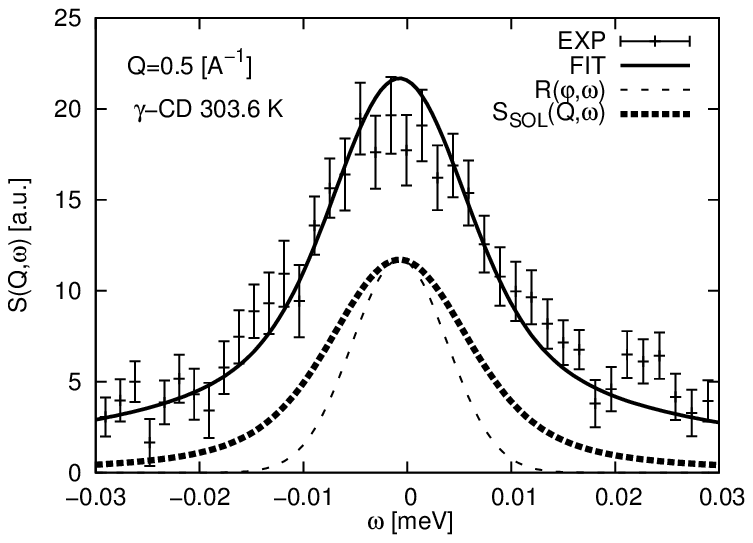}
\hfill
\includegraphics[width=0.45\textwidth]{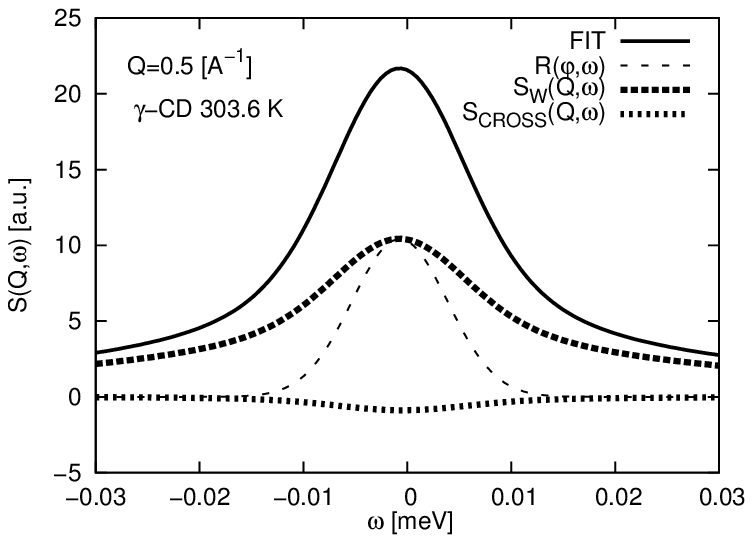}\\
\includegraphics[width=0.45\textwidth]{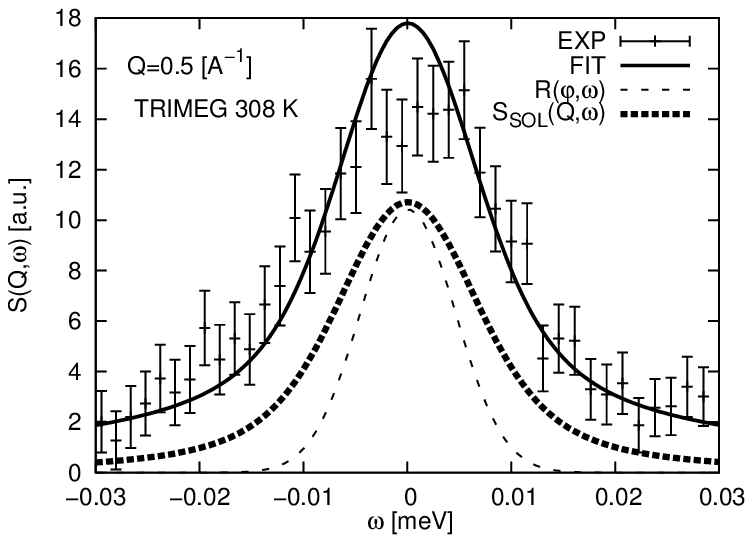}
\hfill
\includegraphics[width=0.45\textwidth]{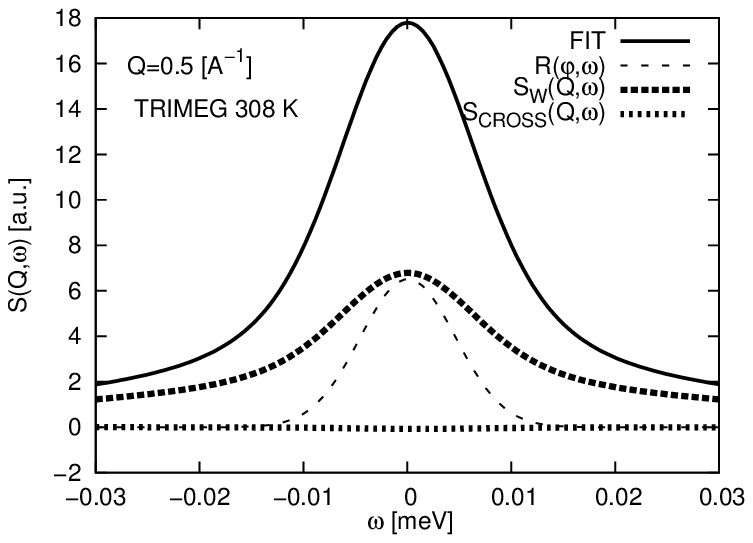}\\
\includegraphics[width=0.45\textwidth]{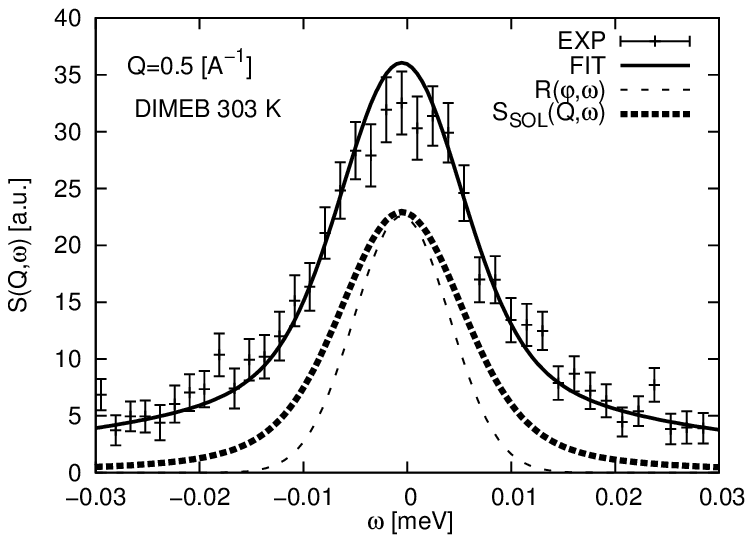}
\hfill
\includegraphics[width=0.45\textwidth]{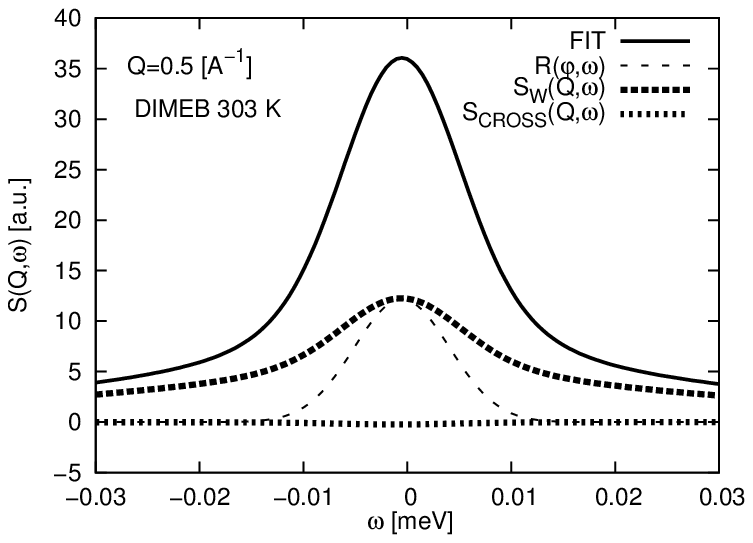}\\
\caption{Examples of fitting of the model to the QENS spectra of cyclodextrins dissolved in D$_2$O, for the experiment with  $\lambda _0$=10 \AA. The notations are the same as in Fig.~\ref{fig:fitlowq}. As compared to Fig.~\ref{fig:fitlowq}, \emph{i}) the broadening of the $S_{sol}(Q, \omega)$ and $S_w(Q, \omega)$-terms is clearly observable; \emph{ii}) the intensity of the $S_{cross}(Q, \omega)$-term is negligible. The scattering is due to $S_{sol}(Q, \omega)$ and due to the scattering from D$_2$O that is practically the same as the pure D$_2$O scattering. }\label{fig:fithigherq} 
\end{figure}

Examples of the fit results  are shown in Fig.~\ref{fig:fitlowq} for the elastic wave vector  
transfer $Q$=0.14 \AA \textsuperscript{-1}, and in Fig.~\ref{fig:fithigherq} for $Q$=0.5 \AA \textsuperscript{-1}. To see if there
is any observable broadening at all, the widths of the separately
plotted components of Eq.~(\ref{eq:sqwaqueoussolution_general}) should be compared to the width of the
energy resolution function. In Fig.~\ref{fig:fitlowq}, both $S_{sol}(Q,\omega )$ and
$S_{w}(Q,\omega)$ have widths similar to the resolution width (we had to
scale down $R(\phi, \omega )$ so that the curves would not
entirely overlap). The $S_{cross}(Q,\omega )$-width is the same as that
of $S_{tr~sol-w}(Q, \omega )$ (see Eq.~(\ref{eq:sqwcross_uniform})), and thus, because at low $Q$
the rotational QENS contribution is negligible, is similar to that of
$S_{sol}(Q,\omega )$. In Fig.~\ref{fig:fithigherq}, at a higher $Q$-value, the widths of
both $S_{sol}(Q,\omega )$ and $S_{w}(Q,\omega )$ are clearly greater than
the resolution width, and the $S_{cross}(Q,\omega )$-term has a
negligible intensity.

The broadening of the quasielastic peak due to translational diffusion, taken as
the full width at half maximum (FWHM) can be calculated from Eq.~(\ref{eq:specialsqwsolute})
and Eq.~(\ref{eq:appAfqtrans}) for the solute and water molecules, respectively (multiply
FWHM [ps$^{-1}$] by 0.6583 to convert it to meV units). At $Q$=0.14 \AA \textsuperscript{-1} and 301
K, FWHMs for $\gamma $-CD and D$_2$O are 0.63 $\mu$eV and 5.1 $\mu$eV,
respectively (see Tabs.~\ref{tab:tabled2o} and  \ref{tab:tablecds} for the input values). Because the
resolution width is ${\approx}$10 $\mu$eV, it is clear why the
$S_{sol}(Q,\omega )$-broadening can hardly be seen in Fig.~\ref{fig:fitlowq}. Also not
seen is the broad contribution due to D$_2$O dynamics (with FWHM of 5.1
$\mu$eV) in the $S_{w}(Q,\omega )$ shown in Fig.~\ref{fig:fitlowq}, because at low $Q$ values
the greatest fraction of the $S_{w}(Q,\omega )$-intensity is its coherent
part which has the width of the $S_{sol}(Q,\omega )$-component 
(see Eqs.~(\ref{eq:sqwinterapproximated}), (\ref{eq:sqwwater_uniform1})-(\ref{eq:sqwwater_uniform3})). At higher $Q$, the $S_{w}(Q,\omega)$-broadening is greater than that of $S_{sol}(Q,\omega )$, see Fig.~\ref{fig:fithigherq},
because the effect of the excluded volume becomes negligible, and the
intramolecular scattering from D$_2$O dominates $S_{w}(Q,\omega )$.

The strong decrease of the intensity with increasing $Q$, both for
$S_{w}(Q,\omega )$ and for $S_{cross}(Q,\omega )$, 
(compare Figs.~\ref{fig:fitlowq} and \ref{fig:fithigherq})
is due to a steep decrease of the effective scattering length, $b(Q)$,
for water and solute molecules. The negative sign of $S_{cross}(Q,\omega
)$ is, technically, the consequence of the definition of the number
density by Eq.~(\ref{eq:appD:nwater}). Simply put, this is because the solute molecules
are dispersed not in vacuum, but in a medium with a non zero neutron
coherent scattering length, and this leads to a destructive
interference.

\begin{figure}
  \includegraphics[width=8cm]{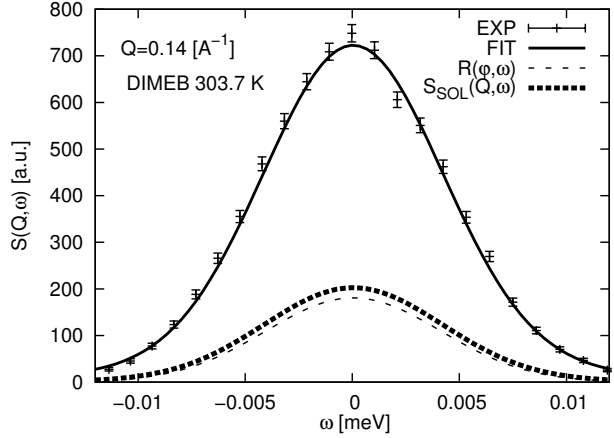}%
  \caption{\label{fig:dimebextra}Fit of the same model to the same spectrum as shown in Fig.~\ref{fig:fitlowq} for DIMEB, the only difference is that the intermolecular D$_2$O scattering due to a finite size of  water molecules was not approximated by the corresponding term for pure D$_2$O (i.e., Eq.~(\ref{eq:sqw_water_2}) was used instead of Eq.~(\ref{eq:sqwinterapproximated})). The notations are the same as in Fig.~\ref{fig:fitlowq}. For reasons why the fit quality is somewhat better, see text. }%
\end{figure}

Since the uniform fluid approximation does not account for the
intermolecular D$_2$O scattering arising due to a finite size of the
molecules, we approximated it by the corresponding contribution to the
pure D$_2$O scattering ($S_{w}^{corr}(Q,\omega )$ in Eq.~(\ref{eq:sqwinterapproximated})). In the low
$Q$ region, this approximation improved the fit quality for $\gamma $-CD, and
slightly worsened the fit quality for TRIMEG and DIMEB (as opposed to
the fits with neglecting $S_{w}^{corr}(Q,\omega )$ entirely, see the example
of such a fit for DIMEB in Fig.~\ref{fig:dimebextra}). Since $S_{w}^{corr}(Q,\omega )$ has
about the same intensity as the D$_2$O-incoherent scattering 
(see section \ref{sec:experimentB}), it can be neglected whenever the total scattering is much more
intense than the D$_2$O-incoherent scattering. As seen from Fig.~\ref{fig:fitlowq}, this
is the case for DIMEB and TRIMEG, but not for $\gamma $-CD. Thus, the
use of $S_{w}^{corr}(Q,\omega )$ is expected to improve the fit quality to a
lesser extent for mCDs than for $\gamma $-CD. The reasons for a
slightly better fit quality for mCDs when the intermolecular D$_2$O
scattering is neglected altogether are difficult to pursue as we can
not at present calculate or measure the exact value of
$S_{w}^{corr}(Q,\omega )$.

At low $Q$ the coherent QENS intensity is
proportional to the squares of the contrast and of the excluded volume
(Eq.~(\ref{eq:lowqsqwtotal3})). The scattering contrasts [10\textsuperscript{-12} cm/\AA \textsuperscript{3}] for solutions in
D$_2$O and the solute molecule's excluded volumes [\AA \textsuperscript{3}]
are: -0.0249 and 1302 ($\gamma $-CD), -0.0525 and 2125 (TRIMEG),
-0.0455 and 1575 (DIMEB) \cite{KUSMINSAENGER2008}. The low contrast for $\gamma $-CD is the
main reason why at low $Q$ the QENS intensity of the $\gamma $-CD spectra
is substantially weaker than that of the mCDs spectra.

As shown above, the quasielastic broadening due to the translational
diffusion of a solute molecule is about 5\% of the resolution width at
$Q$=0.14 \AA \textsuperscript{-1}. This broadening quickly rises with increasing $Q$, thus
allowing to determine $D_{tr~sol}$ by fitting. The so-obtained $D_{tr~
sol}$ values depend on the value at which the rotational diffusion
coefficient, $D_{r~sol}$, was fixed (see Table \ref{tab:tablecds}). This is not surprising:
the radius of a cyclodextrin molecule is about 10 \AA, therefore, the
rotational broadening is non negligible already at $Q$=0.2 \AA \textsuperscript{-1} (i.e., 
for $l > 0$, the coefficients $A^{l}(Q)$ in Eq.~(\ref{eq:appA:AlQcoef}) are
comparable to or greater than $A^{0}(Q)$). Thus, the more we increase $D_{r~
sol}$, the smaller $D_{tr~sol}$ values we get. For DIMEB and TRIMEB the $D_{tr~
sol}$ values obtained with $D_{r~sol}$=0 are fairly close to the ones from
the literature, while for $\gamma $-CD they differ substantially. This
may have to do with a smaller QESANS intensity from $\gamma
$-CD-solutions, or a greater rotational diffusion coefficient. However,
the comparison of the $D_{tr~sol}$ values from different sources must be 
made with caution. Since at low $Q$ most of the scattering is coherent,
the fitted $D_{tr~sol}$-value will depend on how the solute
intermolecular structure factor and hydrodynamic interactions were
taken into account (see, e.g., Ref.~\onlinecite{LONGEVILLEKALI2003}), while from a PFG-NMR experiment a
 true self-diffusion coefficient is obtained. For DIMEB at $\approx$ 303 K the $D_{tr~sol}$-values obtained for $\lambda _0$=10 \AA{}  are
 substantially higher than for $\lambda _0$=15.3 \AA. 
This is in accord with a greater weight of the low $Q$ region for $\lambda _0$=15.3 \AA: at smaller $Q$ a greater fraction of the scattering is coherent, and  therefore  the weight of the collective diffusion coefficient is greater, too.  %

The fitted $D_{r~sol}$ values (with $D_{tr~sol}$ kept fixed) are in a
qualitative agreement with the available data (from Ref.~\onlinecite{SALVADORICD1993} or from the 
Debye-Stokes-Einstein relation for a sphere: $D_r = k_B T/ 6 \eta V_{sphere}$).
Having a wider $Q$ range or a higher energy resolution or both should
help to determine $D_{r~sol}$ with a better precision; this could also allow
the simultaneous determination of $D_{tr~sol}$- and $D_{tr~sol}$ values.%

\begin{figure}
  \includegraphics[width=8cm]{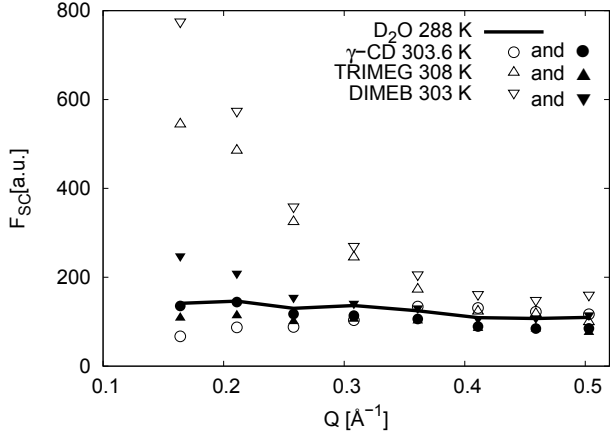}%
  \caption{\label{fig:scalingfactor} Curves of the experimental scaling factor  $F_{sc}(\phi)$ obtained in fitting of QENS spectra from the experiment carried out with $\lambda _0$=10 \AA. The approximately correct $F_{sc}(\phi)$-curve (which should ideally be a horizontal straight line)  is given by $F_{sc}(\phi)$ of D$_2$O. The filled and empty symbols give the $F_{sc}(\phi)$ values obtained with and without detailed consideration of intermolecular D$_2$O-D$_2$O and D$_2$O-solute coherent scattering, respectively. This corresponds to using Eq.~(\ref{eq:sqwaqueoussolution_general}) and Eq.~(\ref{eq:sqwprevious}), respectively.
 }%
\end{figure}

The fits shown in Figs.~\ref{fig:fitlowq}, \ref{fig:fithigherq}, \ref{fig:dimebextra} are satisfactory; however,
approximately the same fit quality could be obtained with the model we
used previously \cite {KUSMINLECHNERSAENGER2007,KUSMINPHDTHESIS}:
\begin{equation}
  S(Q, \omega) = S_{sol}(Q, \omega ) + S_{w}(Q, \omega )
  \label{eq:sqwprevious}
\end{equation}
where water scattering was calculated from Eqs.~(\ref{eq:sqw_water_1})-(\ref{eq:sqw_water_3}) using the structure
factor of pure D$_2$O, and solute scattering was calculated just as it was
done here. Moreover, the solute translational diffusion coefficients
that were obtained were similar to the values obtained in this work.
The reason why Eq.~(\ref{eq:sqwprevious}) ``worked'' is the following: at low $Q$ values the uniform fluid model is a good approximation   and thus the QENS line shape is governed by
the solute dynamics alone. However, with Eq.~(\ref{eq:sqwprevious}), the fitted
$F_{sc}(\phi )$ values for DIMEB and TRIMEG (Fig.~\ref{fig:scalingfactor}, open symbols) are
up to 8 times higher (at low $Q$) than they should be (as suggested by
the curve of the experimental scaling factor  $F_{sc}(\phi )$ - see Eq.~(\ref{eq:sqfit}) - for pure D$_2$O, Fig.~\ref{fig:scalingfactor}). Fitting of the model
developed in this work results in the reasonable $F_{sc}(\phi )$-curves
for $\gamma $-CD and TRIMEG (Fig.~\ref{fig:scalingfactor}, filled symbols). As for DIMEB,
$S_{cm~sol}(Q)$ that we used accounts for hard body solute-solute
interactions only, and solute-solute interactions in DIMEB solutions
are substantially attractive \cite{KUSMINSAENGER2008}. In fact, the excess in the $F_{sc}(\phi
)$ of DIMEB, (increasing toward low $Q$), is in semi-quantitative
agreement with experimental $S_{cm~sol}(Q)$ data \cite{KUSMINSAENGER2008}. Thus, the model developed in the present 
paper provides a good description not only for the line shape, but for the
intensity of the QENS spectra as well. 
\section{\label{sec:discussion}Discussion}
%
%
% DISCUSSION
%
%
The basic goal of this paper is to develop  model expressions  
allowing  an explicit consideration of
 the QENS contributions due to time-dependent spatial pair
correlations between all atoms in aqueous solutions of one molecular species. In section \ref{sec:theoryA} we
showed how one could realize this in general. In section \ref{sec:theoryB}  we applied
an extension of Soper's theory of the excluded volume effect to 
derive the simplified QENS model expressions valid in the low
$Q$ region. In section \ref{sec:theoryC} these model expressions were shown to be
compatible with the concept of scattering contrast. Finally, we
demonstrated that our model can adequately explain the QENS spectra of
cyclodextrins dissolved in heavy water.

An adequate description, in our opinion, comprises not only an
adequate fitting quality (that is, a sufficiently good  line shape description), but an
adequate reproduction of QENS intensities as well. 
As shown above, a criterion for the latter could be the scattering-angle dependence  
which results from the fit of the scaling factor $F_{sc}(\phi )$. 
%The criterion for the latter can serve, as shown above, a $\phi$(scattering angle)-dependence of the factor 
% that scales the calculated scattering function to the experimental data, $F_{SC}(\phi )$. 
%
Whenever this
scaling factor is strongly $\phi $-dependent this means that the
coherent scattering from the sample was not accounted for properly.
Even if one is interested in the dynamics only, e.g., in the rotational
and translational diffusion coefficients, the correct coherent
scattering intensity matters, because this intensity gives the weight
of collective diffusion  vs. self-diffusion, and the weight
of the purely translational QENS component relative to the components
that also contain rotational broadening. Thus, a wrong evaluation of
the coherent scattering intensity results in wrong $D_{tr~sol}$- and $D_{r~
sol}$-values. At somewhat larger  $Q$, when  diffusive  translational water
dynamics becomes increasingly important, although  the  coherent scattering
contribution is not negligible, an incorrect evaluation of the
coherent scattering distorts the mutual proportion of the scattering
due to water and solute dynamics. This leads to wrong   values of dynamical
parameters, even when the fit quality is good. Note that the
determination of $D_{tr~sol}$- and $D_{r~sol}$-values from the QENS spectra
at higher $Q$  (when incoherent scattering dominates) is far
less trivial because in this case intramolecular solute (and solvent)
dynamics contributes to the overall broadening to a much larger
extent.  

  We assumed that the differences  \emph{i}) between dynamical parameters of a single bulk water
molecule in solution and in pure water and  \emph{ii}) between the structure and dynamics of hydration water and bulk water,
are negligible. The first
assumption is justified for such dilute solutions as used here (solute
volume fraction {\textless} 0.05) but is not required. Instead of
keeping $D_{tr~w}$, $D_{r~w}$, etc fixed to the corresponding values found for
pure D$_2$O or H$_2$O, we could adjust them, or even make them free fitting
parameters. In fact, we tried to determine $D_{tr~w}$ in our solutions by
fitting; the obtained values differed from those for pure D$_2$O to a
negligible extent. The second assumption (previously discussed in section \ref{sec:theoryA})
 is fully justified for dilute solutions, because only a small fraction of
water molecules belongs to the hydration shell, and because any change (relative to the
bulk water) in the single molecule dynamics of hydration water 
 has a small contribution  to $S_{w}(Q,\omega )$ in the low $Q$ region. With increasing 
 solute concentration the fraction of hydration water increases, and both assumptions mentioned above become inapplicable. In case of crowded solutions, however, there is no bulk water at all; hence, one set of dynamical parameters (different from such for pure water) may be sufficient to satisfactorily describe  $S_{w}(Q,\omega )$.

Using Vineyard's convolution approximation to calculate 
the scattering function for the collective translational diffusion has little theoretical foundation 
and is known to fail at very low  $Q$ (in the hydrodynamic limit), and in the high $Q$-region \cite{TassoSpringerQENS1972CONVAPP} (the start of which is approximately given by the position of the first peak of the intermolecular structure factor). We stress that this approximation is used for practical purposes only.  
Furthermore, for cyclodextrin solutions, the high  $Q$-region starts from 0.4-0.5 \AA \textsuperscript{-1} (Ref. \onlinecite{KUSMINSAENGER2008}) and  the $Q$-values are not very low, either; hence, the convolution approximation may be acceptable.

With the model presented in section \ref{sec:theoryB} it is straightforward to
obtain the translational and rotational diffusion coefficients of the
solute molecules from quasielastic small-angle neutron scattering
(QESANS) experiments. The obvious requirement is a sufficiently high
energy resolution (i.e.,  sufficiently narrow FWHM of 
$R(\phi ,\omega )$), in order to observe the translational or rotational
broadening, or both. For cyclodextrins, the
resolution used in this work ($\Delta E{\approx}$10 $\mu$eV) is already high
enough, but additional measurements with  the resolution of  the backscattering spectrometers ($\Delta
E{\approx}$1 $\mu$eV) would be rather beneficial. To profit from the simplicity
associated with the uniform fluid approximation, the scattering
contrast should be high, and the measurements should be done in the
 $Q$ region where the QESANS intensity dominates. 
%%%%

It is important to note that the incoherent scattering by the solvent is
 less of a problem in QESANS than it is in SANS, because the broadening
due to solvent dynamics is larger than that of the QESANS component,
and a clear separation between both (given a sufficient energy
resolution) is easy. Thus, even QESANS measurements of H$_2$O solutions 
(despite a high incoherent scattering contribution) would be  
perfectly feasible. The incoherent solute scattering is obviously not a
problem either, except for the fact that it depends on the
translational self diffusion, while the lineshape of the QESANS
component is governed by the collective translational diffusion.
%%%
\subsection*{\label{sec:discussion2}Beyond the uniform fluid approximation}
The uniform fluid approximation is a convenient way to study solute
dynamics without the need to bother with water-water ($S_{cm~w}(Q)$) and
solute-water ($S_{sol-w}(Q)$) structure factors 
(to obtain the solute-solute
structure factor ($S_{cm~sol}(Q)$) is relatively easy). 
On the other hand,
this approximation is limited to the region of low $Q$ values, and it
does not allow to extract any information on the motion of water
molecules relative to the solute molecules (i.e., $I_{tr~sol-w}(Q, t)$ or $S_{tr~
sol-w}(Q, \omega )$). This approximation is not strictly necessary: the
framework presented in section \ref{sec:theoryA} is fairly general. However, to use
this framework in the QENS analysis, one would require to model (or
measure) both $S_{cm~w}(Q)$ and $S_{sol-w}(Q)$. 
%%%%%

As an example of what can be learned, consider the hydration shell:
it is the layer where significant time-dependent spatial correlations
between the solute and water molecules exist, and these correlations
contribute to the intermolecular coherent scattering ($S_{cross}(Q,\omega)$ 
in Eq.~(\ref{eq:sqwaqueoussolution_general})). Thus,  as seen from Eq.~(\ref{eq:sqwcrossterm}),  
to learn about the hydration shell we need an expression for $S_{tr~sol-w}(Q, \omega )$. 
This expression can be taken from Eq.~(\ref{eq:convapp_solw}), but 
an even more  simple approach (which does not rely on Vineyard's approximation) could be to write: 
\begin{equation}
  S_{tr~sol-w}(Q, \omega ) = S_{sol-w}(Q) \text{Lor}(D_{tr~sol-w}Q^2, \omega )  
  \label{eq:tentativehydration}
\end{equation}
where $S_{sol-w}(Q)$ is given by Eq.~(\ref{eq:appC}).
Here the $Q^2$-dependent width of the conjectured Lorentzian is controlled by an apparent collective diffusion coefficient, 
 $D_{tr~sol-w}$, 
 which would be analogous to collective diffusion coefficients defined for one molecular species, 
but would originate exclusively from the diffusive motion of solute and solvent molecules relative to each other. 
This coefficient will depend on the strength of the interactions between the two different molecules, 
just as for instance in the case of solute-solute collective diffusion (see Refs.~\onlinecite{LONGEVILLEKALI2003,LechnerLongevilleChapter}).
Therefore the $D_{tr~sol-w}$-value could serve
as a measure of solute-water interactions and would be related to the time spent by a
water molecule in the hydration shell. If $S_{cm~w}(Q)$ and $S_{sol-w}(Q)$ were
known or could be modeled, then, given an energy resolution is
sufficiently high to see the change of the QENS lineshape due to
$S_{cross}(Q,\omega )$, the experimental determination of $D_{tr~sol-w}$ would
be possible.
%%%%%

At present it is not easy to obtain $S_{cm~w}(Q)$ and $S_{sol-w}(Q)$ from a
QENS experiment. Indeed, the QENS lineshape depends on the
$Q$-dependent intensities of three separate components: $S_{sol}(Q,\omega
)$, $S_{cross}(Q,\omega )$ and $S_{w}(Q,\omega )$; these intensities depend
on the structure factors $S_{cm~sol}(Q)$, $S_{sol-w}(Q)$, and $S_{cm~w}(Q)$. If the
scaling factor, $F_{sc}(\phi )$, would result from the fit as  an angle-independent constant 
(as expected in the error-free ideal case),
both $S_{sol-w}(Q)$, and $S_{cm~w}(Q)$ could be obtained from fitting the model
to the QENS spectra, and then compared to theoretical models. If this
were true in the present work, then $F_{sc}(\phi )$ would be $\phi
$-independent for the case of pure D$_2$O but, as seen in Fig.~\ref{fig:scalingfactor}, the
corresponding $F_{sc}(\phi )$-curve is still not entirely flat. This is
mainly because of the multiple scattering and the  
attenuation of the incident and singly scattered beams in the sample 
 (in case when the sample container is a plain slab the attenuation is especially $\phi $-dependent). 
 The corrections for these two effects 
 depend on a number of
different factors (sample size, macroscopic scattering and absorption
cross-sections of the sample, scattering angle, sample orientation etc)
and, to be exact, require a numerical integration of multiple
integrals. If the sample container is a hollow cylinder, the
 attenuation of single scattering is independent on the scattering angle and the multiple scattering is less important; hence, 
 a flatter $F_{sc}(\phi )$-curve can be expected. Then, one can keep
$F_{sc}$ as a $\phi $ -independent fit parameter and obtain
$Q$-dependent structure factors from the fit to the QENS spectra.

In conclusion, we presented a model accounting for both coherent and
incoherent quasielastic neutron scattering from an aqueous solution,
and demonstrated how this model together with an extension of
Soper's theory of the excluded volume effect\cite{SOPER1997EXCLUDEDV} to the
time-dependent case can reproduce the experimental QENS spectral
lineshapes and intensities. The model explained the quasielastic
small-angle neutron scattering spectra of D$_2$O solutions of
cyclodextrins without an \emph{ad hoc} assumption on the scattering by the
hydration shell made earlier\cite {KUSMINLECHNERSAENGER2007,KUSMINPHDTHESIS}. While the model potentially allows the
characterization of the hydration shell, this was not possible with our
present QENS spectra. But this may be feasible in the future with
better measurement conditions: better statistics, higher energy resolution, and if the
QESANS experiment is performed with much stricter observation of
conditions of accuracy at low scattering angles, than has been
standard in the past mainly in large-angle scattering experiments.
This accuracy requirement concerns the precision of our knowledge of the sample container geometry, 
sample size, scattering angles, sample orientation etc. 
\begin{acknowledgments}
  % put your acknowledgments here.
REL thanks G. Steiner and B. Urban for very valuable technical
assistance during the NEAT experiment. Two of us (AK and REL) thank the
Hahn-Meitner-Institut for its hospitality. Financial support for AK by the
Institute of Chemistry and Biochemistry of the Freie Universit\"at  Universität Berlin
is gratefully acknowledged.
\end{acknowledgments}
%
% Specify following sections are appendices. Use \appendix* if there
% only one appendix.
%
% MULTI PART APPENDIX
%
\appendix
%
% MULTI PART APPENDIX
%
% SCATTERING BY ONE MOLECULAR SPECIES
%
\section*{Frequently used symbols}
\small{
 \flushleft{ $A^{l~coh}_w(Q)$ and $A^{l~inc}_w(Q)$ ($A^{l~coh}_{sol}(Q)$ and $A^{l~inc}_{sol}(Q)$) -- coefficients in Sears's expansion 
 of the rotational correlation function of a water (solute) molecule.}\\
 $A^{l~(p)}(Q)$  -- coefficients in Sears's expansion 
 of the rotational correlation function of a solute excluded volume.\\
 $b_w(Q)$ ($b_{sol}(Q)$) -- the effective scattering length of a water (solute) molecule.\\
 %
% $G_{tr~w}(r,t)$ and $G_{tr~sol-w}(r,t)$ (or $G_{w}(r,t)$ and $G_{sol-w}(r,t)$) -- time-dependent translational water-water and solute-water pair correlation functions \\
 $G_{w}(r,t)$, $G_{sol}(r,t)$ and $G_{sol-w}(r,t)$ -- time-dependent translational water-water, solute-solute, 
 and solute-water pair correlation functions.\\
$G_{sol}^{dist(p)}(r, t)$ ($G_{sol}^{self (p)}(r, t)$)  --  the orientational  correlation function for volume
elements  which belong to two distinct solute molecules (the same solute molecule).\\
 $N^{(p)}(Q)$  -- an analog of $b(Q)$ for a solute excluded volume.\\
 $N_w$ ($N_{sol}$) -- the number of water (solute) molecules in  solution.\\
 $n_w$ ($n_{sol}$) -- the number density of water (solute) molecules in  solution.\\
 $\nu ^{(p)}$ ($V^{(p)}$) -- the volume excluded by a single (all) solute molecule(s) in solution.\\
%
%$S_w(\bm{Q},\omega)$, $S_{sol}(\bm{Q},\omega)$, $S_{cross}(\bm{Q},\omega)$ -- 
$S_w(Q,\omega)$, $S_{sol}(Q,\omega)$, $S_{cross}(Q,\omega)$ -- 
% ($I_w(\bm{Q},t)$, $I_{sol}(\bm{Q},t)$, $I_{cross}(\bm{Q},t)$) (intermediate scattering)
 scattering  functions originating from  water-water, solute-solute and solute-water pair-correlations, respectively.\\
$S_{cm~sol}(Q)$, $S_{cm~w}(Q)$, and $S_{sol-w}(Q)$  -- solute-solute, water-water, and solute-water intermolecular CM structure factors in
solution.\\
 $S^{l~coh}_w(Q,\omega)$ ($S^{l~coh}_{sol}(Q,\omega)$) -- the $l$th scattering function component in Sears's expansion 
 of the rotational correlation function of a water (solute) molecule.\\
 $S^{l~(p)}_w(Q,\omega)$  -- the $l$th scattering function component in Sears's expansion 
 of the rotational correlation function  of a solute excluded volume.\\
  $S^{coh}_{tr~w}(Q, \omega )$ and $S^{inc}_{tr~w}(Q, \omega )$  ($S^{coh}_{tr~sol}(Q, \omega )$ and $S^{inc}_{tr~sol}(Q, \omega )$) -- the coherent and incoherent  translational
scattering functions for the CM of water (solute) molecules.\\
$S_{tr~sol-w}(Q, \omega )$ -- the (coherent) scattering function for the translational motion of water molecules relative to the solute molecules.\\
  %

%\begin{table}
%\caption{\label{tab:scatfunctions} }
%\begin{tabular}{lccc}
%  \hline
%  Space-Time &
%  Reciprocal space-Time &
%  Reciprocal space-Frequency &
%
%  \hline
%\end{tabular}
%\end{table}

}

\section{\label{sec:appendixA}QENS model for one molecular species}
%
% SCATTERING BY ONE MOLECULAR SPECIES
%
In section \ref{sec:theoryA} the scattering by solute and water molecules, 
$S_{sol}(Q, \omega)$  and $S_w(Q, \omega )$, respectively, was expressed using a model developed by Sears \cite{SEARS1967P3}. 
In the following we show the deduction of this result in more detail.

The scattering function for molecules of one particular species in a
liquid solution is,
\begin{equation}
S(Q,\omega)=n \text{DWF}\sum _{l=0}^{\infty }S^{l}(Q,\omega ) 
  \label{eq:appA:total}
\end{equation}
where $n$ is the number density of the molecules. The Debye-Waller
factor, DWF = $e^{- \left < u^2 \right > Q^2}$,
accounts for the $Q$-dependent decrease (caused by vibrational motions)
of the quasielastic intensity, $\left < u^2 \right >$ is the
mean square vibrational amplitude of a molecule. In writing of Eq.~(\ref{eq:appA:total})
we used the model of continuous rotational diffusion on a spherical
surface \cite{SEARS1967P3}, thus:
\begin{equation}
  S^{0}(Q,\omega)=A^{0~coh}(Q)S^{coh}_{tr}(Q,\omega)+
A^{0~inc}(Q)S_{tr}^{inc}(Q,\omega )\quad l = 0  
  \label{eq:appA:totall0}
\end{equation}
\begin{equation}
S^{l}(Q,\omega)=(2l+1) A^{l}(Q) S_{tr}^{inc}(Q,\omega
)\otimes \text{Lor}(l(l+1)D_{r},\omega ) \quad l \neq 0
  \label{eq:appA:totallall}
\end{equation}
where $D_{r}$ is the rotational diffusion coefficient of the molecule. 
% here I deleted the relation between Dr and tauR, see a MS file with a correction on 18-Feb-2010.
The coefficients $A^{l}(Q)$ account for the
molecule's coherent and incoherent scattering and are given by:
\begin{equation}
A^{l}(Q) = A^{l~coh}(Q) + A^{l~inc}(Q)= \sum _{\mu , \nu=1}^{m, m}{
   [  \left < b_{\mu} \right > \left < b_{\nu} \right > + \frac{\sigma _{\mu}^{inc}\delta _{\mu \nu }}{4\pi }] 
    j_{l}(Qr_{\mu })j_{l}(Qr_{\nu })P_{l}(\cos \theta _{\mu \nu })       }
  \label{eq:appA:AlQcoef}
\end{equation}
where $m$ is the number of nuclei in the molecule,  $\left < b _{\mu} \right > $
 is the neutron coherent scattering length of the $\mu $th nucleus, 
 the vectors $ \bm{r}_\mu $ and $\bm{r}_\nu $ point from the CM to the $\mu $th and
$\nu$th atoms, $\theta _{\mu \nu} $ is the angle between $ \bm{r}_\mu $ and $\bm{r}_\nu$, $P_l$
is the Legendre polynomial of degree $l$, $\sigma _{inc}$ is the incoherent
scattering cross-section. Note that for $l \neq 0$, because of the assumption that rotational motions 
of different molecules are not correlated with each other \cite{SEARS1967P3},
only 
% the incoherent translational scattering function,
$S_{tr}^{inc}(Q,\omega)$ appears in Eq.~(\ref{eq:appA:totallall}).

For $l = 0 $, in Vineyard's  convolution approximation \cite{VINEYARD1958},
\begin{equation}
S^{0}(Q,\omega)=\left ( A^{0~coh}(Q)(S_{cm}(Q)-1)+A^{0~inc}(Q) \right ) S_{tr}^{inc}(Q,\omega )  
  \label{eq:appA:sqw0convolutionapplied}
\end{equation}
The function $S_{cm}(Q)$ is the intermolecular center-of-mass (CM)
structure factor of the molecules.

The incoherent translational scattering function, $S^{inc}_{tr}(Q,\omega)$,
is a Lorentzian:
\begin{equation}
S_{tr}^{inc}(Q,\omega)=\frac{1}{\pi }\frac{f_{tr}}{f_{tr}^{2}(Q)+\omega
^{2}}=\text{Lor}(f_{tr}(Q),\omega)  
  \label{eq:appA:sqwtrinc}
\end{equation}
In the frame of the isotropic jump-diffusion model \cite{EgelstaffLiquidState1967}:
\begin{equation}
  f_{tr}(Q) = D_{tr}Q^2/(1 + \tau _{tr}D_{tr}Q^2) 
  \label{eq:appAfqtrans}
\end{equation}
where $D_{tr}$ and $\tau _{tr}$ are the molecule's
translational diffusion coefficient and correlation time, respectively.

We define the effective scattering length of the molecule, $b(Q)$:
\begin{equation}
b(Q)=\sum _{\mu =1}^{m}{ \left < b_{\mu } \right > \frac{\sin Qr_{\mu }}{Qr_{\mu }}     }  
  \label{eq:appA:beffQdefine}
\end{equation}
Note that:
\begin{equation}
  A^{0~coh}(Q)=b^{2}(Q)
  \label{eq:appA:beffqa0coh}
\end{equation}

The model defined above is applied in section \ref{sec:theoryA} to express 
the scattering by solute and water molecules, 
$S_{sol}(Q, \omega)$ and $S_w(Q, \omega )$, respectively.

%
% INTERMEDIATE SCATTERING FUNCTION: CROSS TERM
%
\section{\label{sec:appendixB}Intermediate scattering function for solute-water pair-correlations}
%
% INTERMEDIATE SCATTERING FUNCTION: CROSS TERM
%
In the following we show how the term $S_{cross}(Q,\omega)$ appearing in  Eq.~(\ref{eq:sqwaqueoussolution_general})
 leads to Eq.~(\ref{eq:sqwcrossterm}). 
The contribution of water-solute cross-correlations to Eq.~(\ref{eq:iqtaqueoussolution_general}) can be written as
\begin{equation}
I_{cross}(\bm{Q},t)  = 
\sum _{\mu =1}^{} \left < b_{\mu} \right > \left\langle  e^{-i\bm{Qr}_{\mu} }\right\rangle \sum _{\nu =1}^{} \left < b_{\nu}
\right > \left\langle  e^{i\bm{Qr}_{\nu}(t)}\right\rangle I_{tr~cross}(\bm{Q},t)
  \label{eq:appB:iqtcrossgeneral}
\end{equation}
where the summations over $\mu$ and over $\nu$ are taken over the nuclei in
the solute and in the water molecule, respectively, and
$ \left <b_{\mu} \right >$ is the neutron coherent scattering length
of the $\mu$th nucleus. We assumed that \emph{i}) the rotational motions of a
water molecule, as well as of a solute molecule are uncorrelated with
their translational motions; \emph{ii}) the rotational motion of a water
molecule is uncorrelated with the rotational motion of a solute
molecule. The translational contribution, $I_{tr~cross}(\bm{Q},t)$, reads 
\begin{equation}
  I_{tr~cross}(\bm{Q},t)  = \sum _{i=1}^{N_{sol}}  \sum _{j=1}^{N_w}  \left\langle e^
  {-i\bm{Q}(\bm{R}_{i}-\bm{R}_{j}(t))}\right\rangle  + \sum _{i=1}^{N_w} \sum _{j=1}^{N_{sol}}  \left\langle e^{
-i\bm{Q}(\bm{R}_{i}-\bm{R}_{j}(t))}\right\rangle 
  \label{eq:appB:iqtcrosstrans}
\end{equation}
Note that in the first double sum the index $i$ refers to the CM of a
solute molecule and  $j$ to  the CM of a water molecule, while in the second
double sum the order is opposite.

Since $I(\bm{Q},t)$ (and $S(\bm{Q},\omega)$) measured in the experiment are the averages over the measurement time ($t_{m}$), and because 
all solute and water molecules are equivalent,
Eq.~(\ref{eq:appB:iqtcrosstrans}) can be written as
\begin{eqnarray}
%  I_{tr~cross}(\bm{Q},t)  =  \frac{1}{t_{m}-t} \int _{0}^{t_{m}-t} \mathrm{d}t \sum _{i=1}^{N_s}  \sum _{j=1}^{N_w}   \exp
%(-i\bm{Q}(\bm{R}_{i}(t_0)-\bm{R}_{j}(t+t_0)))  + \sum _{i=1}^{N_w} \sum _{j=1}^{N_s}   \exp
%(-i\bm{Q}(\bm{R}_{i}(t_0)-\bm{R}_{j}(t+t_0)))
I_{tr~cross}(\bm{Q},t)  =  \frac{1}{t_{m}-t} \int _{0}^{t_{m}-t} \mathrm{d}t_0 (N_{sol}  \sum _{j=1}^{N_w}  e^ 
{-i\bm{Q}(\bm{R}_{sol}(t_0)-\bm{R}_{j}(t+t_0))}  \nonumber \\
+ N_w \sum _{j=1}^{N_{sol}}   e^{
-i\bm{Q}(\bm{R}_{w}(t_0)-\bm{R}_{j}(t+t_0))} )
  \label{eq:appB:iqtcrosstranstime0}
\end{eqnarray}
Henceforth,  since $t \ll t_{m}$, we approximate $t - t_{m}$ by $t_{m}$. 
By introducing  $G_{sol-w}(\bm{r},t,t_0)$ and $G_{w-sol}(\bm{r},t,t_0)$ which are solute-water and water-solute time-dependent pair-correlation functions, respectively, and by presenting the sums as integrals of these functions,
  Eq.~(\ref{eq:appB:iqtcrosstranstime0}) can be written as

\begin{eqnarray}
  I_{tr~cross}(\bm{Q},t)  =  \frac{1}{t_{m}} \int _{0}^{t_{m}} \mathrm{d}t_0  \int _{V}  
  e^{i\bm{Q}\bm{r}} \left ( N_{sol} G_{sol-w}(\bm{r},t,t_0)   
  + N_w  G_{w-sol}(\bm{r},t,t_0) \right ) \mathrm{d}\bm{r} 
  \label{eq:appB:iqtcrosstranstime0grt}
\end{eqnarray}
where $V$ is the volume of the sample. Note that $G_{sol-w}(\bm{r},t,t_0)$ and $G_{w-sol}(\bm{r},t,t_0)$  are averages over initial positions of the solute and water molecule, respectively.

To introduce the dependence on the spatial origin $\bm{r_0}$ via time-dependent local number
densities \cite{VanHove1954,MarshallLovesey1971}, $n(\bm{r}, t)$ (for the definition see Eqs.~(\ref{eq:appD:numberdensity}-\ref{eq:appD:grtdefinition}) in Appendix \ref{sec:appendixD}), we define
 \begin{equation}
  G_{sol-w}(\bm{r},t,t_0)=\frac{1}{N_{sol}}\int  n_{sol}(\bm{r_0},t_0)n_{w}(\bm{r_0}+\bm{r},t+t_0)  \mathrm{d}\bm{r_0}\\
  \label{eq:crossgrtt0solw}
 \end{equation}
 \begin{equation}
  G_{w-sol}(\bm{r},t,t_0)=\frac{1}{N_{w}}\int  n_{w}(\bm{r_0},t_0)n_{sol}(\bm{r_0}+\bm{r},t+t_0)  \mathrm{d}\bm{r_0}
  \label{eq:crossgrtt0wsol}
 \end{equation}
Eq.~(\ref{eq:appB:iqtcrosstranstime0grt}) can now be written as

\begin{eqnarray}
  I_{tr~cross}(\bm{Q},t)  =  \frac{1}{t_{m}} \int _{0}^{t_{m}} \mathrm{d}t_0  \int _{V}  
   e^{-i\bm{Q}\bm{r}} \mathrm{d}\bm{r} \int _{V} n_{sol}(\bm{r_0},t_0) n_{w}(\bm{r+r_0},t+t_0) \mathrm{d}\bm{r_0} \nonumber  \\
  +  \frac{1}{t_{m}} \int _{0}^{t_{m}} \mathrm{d}t_0 \int _{V}    e^{-i\bm{Q}\bm{r}}  \mathrm{d}\bm{r} \int _{V} 
   n_{w}(\bm{r_0},t_0)  n_{sol}(\bm{r+r_0},t+t_0) \mathrm{d}\bm{r_0} 
  \label{eq:appB:iqtcrosstranstime0r0}
\end{eqnarray}
In principle, Eq.~(\ref{eq:appB:iqtcrosstranstime0r0}) is just an expanded version of Eq.~(\ref{eq:appB:iqtcrosstrans}) with averaging over initial positions and times shown explicitly.
Since the functions $n_w(\bm{r},t)$ and $n_{sol}(\bm{r},t)$ are real-valued, the two terms 
 at the right side of  Eq.~(\ref{eq:appB:iqtcrosstranstime0r0}) are identical. 
 Thus, one can see that $I_{tr~cross}(\bm{Q},t)$, and, consequently, the cross-term $I_{cross}(\bm{Q},t)$ is controlled by the relative motion of a water molecule with respect to a solute molecule, and \emph{vice versa}.
  For the reason given in Appendix~\ref{sec:appendixC}, from the two possible denominations ($G_{sol-w}(\bm{r},t)$  and $G_{w-sol}(\bm{r},t))$  we will use the first one, i.e., 
$G_{sol-w}(\bm{r},t)$ and its Fourier transforms.

From the above, after averaging over all $\bm{Q}$-orientations and using
Eq.~(\ref{eq:appA:beffQdefine}), Eq.~(\ref{eq:appB:iqtcrossgeneral}) can be written as:
\begin{equation}
I_{cross}(Q,t)=2n_{sol}b_{sol}(Q)b_{w}(Q)I_{tr~sol-w}(Q,t)
  \label{eq:appB:icrossfinal}
\end{equation}
where $n_{sol}$ is the solute number density, and $I_{tr~sol-w}(Q,t)$ is the space-Fourier transform of Eq.~(\ref{eq:crossgrtt0solw}). 
The time-Fourier transformation of Eq.~(\ref{eq:appB:icrossfinal}) yields Eq.~(\ref{eq:sqwcrossterm}).

%
% SOLUTE-WATER STRUCTURE FACTOR
%
\section{\label{sec:appendixC}Solute-water  pair-correlations}
In section \ref{sec:theoryA} we related  the scattering contribution from the time-dependent water-solute pair-correlations, 
 $S_{cross}(Q, \omega)$, 
 to $S_{tr~sol-w}(Q, \omega)$, which,  using Vineyard's convolution approximation, was approximated
 by the product of the solute-water structure factor $S_{sol-w}(Q)$ and the water incoherent translational scattering function 
 $S_{tr~w}^{inc}(Q,\omega)$ (see Eq.~(\ref{eq:convapp_solw})). The explanation is as follows.
 As it was said in Appendix \ref{sec:appendixB}, both  
 $S_{tr~sol-w}(Q,\omega)$ and  $S_{tr~w-sol}(Q,\omega )$ can be used. In Vineyard's approximation  one can write
 \begin{equation}
   S_{tr~sol-w}(Q,\omega )=S_{sol-w}(Q)S_{tr~w}^{(sol)~inc}(Q,\omega)  
  \label{eq:convappnew1}
\end{equation}
 \begin{equation}
   S_{tr~w-sol}(Q,\omega )=S_{w-sol}(Q)S_{tr~sol}^{(w)~inc}(Q,\omega)  
  \label{eq:convappnew2}
\end{equation}
In Eq.~(\ref{eq:convappnew1})  $S_{tr~w}^{(sol)~inc}(Q,\omega)$ depends on the self-diffusion of a water molecule
in the coordinate system which has its origin  at a solute molecule. Similarly, in Eq.~(\ref{eq:convappnew2})  
$S_{tr~sol}^{(w)~inc}(Q,\omega)$ depends on the self-diffusion of a solute molecule in the coordinate system with the origin  
at a water molecule. Both equations are correct but neither can   be directly used.
While, in general, $S_{cross}(Q, \omega)$ must depend on both water and solute dynamics, 
since a water molecule diffuses much faster than a cyclodextrin molecule, in the first approximation one could neglect
the translational diffusion of a solute molecule altogether. Thus,
 Eq.~(\ref{eq:convappnew1})  leads to  Eq.~(\ref{eq:convapp_solw}).

%
% SOLUTE-WATER STRUCTURE FACTOR
%
  The solute-water structure factor is:
  \begin{equation}
S_{sol-w}(\bm{Q})=\int e^ {i\bm{Qr}}(G_{sol-w}(\bm{r})-n_{w})\, \mathrm{d}\bm{r}    
    \label{eq:appC}
  \end{equation}
  where by writing $G_{sol-w}(\bm{r}) - n_{w} $ instead of $G_{sol-w}(\bm{r})$ we neglect the
  scattering that can not be observed in practice (at $Q \approx 0$), $n_{w}$
is the mean number density of water in solution. 
For dilute
solutions $G_{sol-w}(\bm{r})$ can be modeled as follows: $G_{sol-w}(\bm{r})$ is $n_{w}$ if water and solute molecules do not
overlap and 0 otherwise. (A similar approach was already
used in QENS analysis (section 2.12 in Ref.~\onlinecite {VASSGILANYI2005})). Eq.~(\ref{eq:appC}) becomes an
integral over the volume which is somewhat larger than the excluded
volume of the solute molecule (to account for the finite size of the
water molecule). Note that $G_{sol-w}(\bm{r})$ does not have to be spherically
symmetric. After averaging over all $\bm{Q}$
orientations, the resulting $S_{sol-w}(Q)$ can be used 
to calculate $S_{tr~sol-w}(Q,\omega )$ in Eq.~(\ref{eq:convapp_solw}).
%
%
% UNIFORM FLUID APPROXIMATION
%
%
%
%  EXTENSION OF SOPER'S THEORY
%
\section{\label{sec:appendixD}Uniform fluid approximation (UFA) in QENS}
%
%  EXTENSION OF SOPER'S THEORY
%
\subsection{\label{subsec:generalformalism}General formalism}
In order to derive the scattering functions given by Eqs.~(\ref{eq:sqwwater_uniform1})-(\ref{eq:sqwwater_uniform3}) 
and Eq.~(\ref{eq:sqwcross_uniform}) in Section \ref{sec:theoryB}, we 
 give here an extension of Soper's (static) theory of the
excluded volume effect\cite{SOPER1997EXCLUDEDV} to the dynamical case   implying time-dependent correlation functions, 
while the static theory obviously is restricted to $t = 0$.   Equations from the
original paper are referred to as Eqs.~(S1), (S2), etc. We abbreviate
terms $\bm{R}_{i}(t=0)$ by $\bm{R}_{i}$, $n(\bm{r}, t = 0)$ by $n(\bm{r})$, and so on.

Instead of the static local number density, $n(\bm{r})$ used in Ref.~\onlinecite{SOPER1997EXCLUDEDV},
the function relevant in our case is the time-dependent local number
density, $n(\bm{r},t)$, which for $N$ atoms in a volume $V$ is:
\begin{equation}
  n(\bm{r},t)=\sum _{j=1}^{N}\delta (\bm{r}-\bm{R}_{j}(t))
  \label{eq:appD:numberdensity}
\end{equation}
where $\bm{R}_{j}(t)$ is the vector giving the position of $j$th atom at time $t$. The
expression for the time-dependent pair correlation function, $G(\bm{r}, t)$,
reads:
\begin{equation}
  G(\bm{r},t)=\frac{1}{N}\int  n(\bm{r'})n(\bm{r'+r},t) \, \mathrm{d}\bm{r'}
  \label{eq:appD:grtdefinition}
\end{equation}
$G(\bm{r}, t)$ can be presented as the sum of the self and distinct
time-dependent correlation functions, $G^{self}(\bm{r},t)$ and $G^{dist}(\bm{r},t)$:

\begin{equation}
G(\bm{r},t)=G^{self}(\bm{r},t)+G^{dist}(\bm{r},t)  
  \label{eq:appD:grt_sum_dist_self}
\end{equation}
\begin{eqnarray}
  G^{self}(\bm{r},t)=\frac{1}{N}\sum _{i=1}^{N}\delta
  (\bm{r}+\bm{R}_{i}-\bm{R}_{i}(t)) \nonumber  \\ G^{dist}(\bm{r},t)=\frac{1}{N}\sum
  _{i\neq j=1}^{N,N}\delta (\bm{r}+\bm{R}_{i}-\bm{R}_{j}(t))
  \label{eq:appD:gself_gdist_define}
\end{eqnarray}

Given that there are $N_{cm}$ molecules, $M$ atoms per molecule, $N = N_{cm}M$, the
functions $n_{cm}(\bm{r},t)$, and $G_{cm}(\bm{r},t)$ are defined as above except that they
refer to the CM of the molecules. Introducing the internal atomic
number density, $n^{(p)}(\bm{r},t)$ (which is zero outside the volume of the
molecule), $n(\bm{r},t)$ can be presented as:
\begin{equation}
  n(\bm{r},t)=\int n_{cm}(\bm{r'},t)n^{(p)}(\bm{r}-\bm{r'},t)\, \mathrm{d}\bm{r'}
  \label{eq:appD:nrtconvolution}
\end{equation}
The $G(\bm{r}, t)$-expression defined by Eq.~(\ref{eq:appD:grtdefinition}) can be rewritten using Eq.~(\ref{eq:appD:nrtconvolution}) as:
\begin{equation}
  G(\bm{r},t)=\frac{1}{N}\int \, \mathrm{d}\bm{r'} \int
  n_{cm}(\bm{r''})n^{(p)}(\bm{r'}-\bm{r''})\, \mathrm{d}\bm{r''}\int
  n_{cm}(\bm{r'''},t)n^{(p)}(\bm{r}+\bm{r'}-\bm{r'''},t)\, \mathrm{d}\bm{r'''}
  \label{eq:appD:grtinternal1}
\end{equation}
Substituting $\bm{u'} = \bm{r'} - \bm{r''}$  and
$ \bm{u''}= \bm{r} + \bm{r'}- \bm{r'''}$ we get:
\begin{equation}
  G(\bm{r},t)=\frac{1}{N}\int \, \mathrm{d}\bm{r'}\int
  n_{cm}(\bm{r'}-\bm{u'})n^{(p)}(\bm{u'})\, \mathrm{d}\bm{u'}\int
  n_{cm}(\bm{r}+\bm{r'}-\bm{u''},t)n^{(p)}(\bm{u''},t)\, \mathrm{d}\bm{u''}
  \label{eq:appD:grtinternal2}
\end{equation}
The substitutions $\bm{u} = \bm{u''}-\bm{u'} $ and $\bm{r'} = \bm{r''} + \bm{u'}$ yield
(compare with Eqs.~(S8, S9)):
\begin{equation}
G(\bm{r},t)=\frac{1}{N}\int \, \mathrm{d}\bm{u}\int \, \mathrm{d}\bm{r''}
n_{cm}(\bm{r''})n_{cm}(\bm{r''}+\bm{r}-\bm{u},t)
\int \, \mathrm{d}\bm{u'}\langle n^{(p)}(\bm{u'})n^{(p)}(\bm{u'}+\bm{u},t)\rangle _{\Omega }
  \label{eq:appD:grtinternal3}
\end{equation}

The self and distinct internal correlation functions, $G^{self(p)}(\bm{u}, t)$
and $G^{dist(p)}(\bm{u}, t)$, respectively, are:
\begin{equation}
G^{self(p)}(\bm{u},t)=\frac{1}{M}\int \langle
n^{(p)}(\bm{u'})n^{(p)}(\bm{u'}+\bm{u},t)\rangle _{\Omega }\, \mathrm{d}\bm{u'} 
  \label{eq:appD:grtselfinternal}
\end{equation}
\begin{equation}
   G^{dist(p)}(\bm{u},t)=\frac{1}{M}\int
   \langle n^{(p)}(\bm{u'})\rangle _{\Omega }\langle n^{(p)}(\bm{u'}+\bm{u},t)\rangle _{\Omega }\, \mathrm{d}\bm{u'} 
  \label{eq:appD:grtdistinternal}
\end{equation}
where $\langle ..\rangle _{\Omega }$ stands for orientational
average. The integrals of $G^{self(p)}(\bm{u}, t)$ and $G^{dist(p)}(\bm{u}, t)$ over the
volume of the molecule are equal to 1 and $M- 1$, respectively. It
follows from the above:
\begin{equation}
G(\bm{r},t)=\int G^{self(p)}(\bm{u},t)G_{cm}^{self}(\bm{r}-\bm{u},t) \, \mathrm{d}\bm{u}   +  
 \int G^{dist(p)}(\bm{u},t)G_{cm}^{dist}(\bm{r}-\bm{u},t) \, \mathrm{d}\bm{u} 
  \label{eq:appD:grtinternalsum}
\end{equation}
For $t = 0$, Eq.~(\ref{eq:appD:grtinternalsum}) is identical to Eq.~(S10).
%
%  APPLICATION FOR WATER-WATER and SOLVENT-WATER
%
\subsection{\label{subsec:uniformapplication}Application of the UFA to water-water and solute-water pair-correlations}
%
%  APPLICATION FOR WATER-WATER and SOLVENT-WATER
%
Let us have $N_{sol}$ solute molecules in a volume $V$, the mean solute
number density is $n_{sol}$, $n_{sol} = N_{sol}/V$. In solution each solute molecule
excludes a volume $\nu^{(p)}$, called the excluded volume in the following; the
total excluded volume is $V^{(p)}$, $V^{(p)} = \nu^{(p)} N_{sol}$. The water number
densities in pure water and solution are $n_0$ and $n_w$, respectively ($n_w =
N_w/V = n_0(V - V^{(p)})/V$). The number of water molecules excluded by one solute molecule is $M$, $M = \nu^{(p)} n_0$. 
 Let us express $N_w$ as:
\begin{equation}
N_{w}=n_{w}V=n_{0}(1-\frac{V^{(p)}}{V})V=M N_{sol}[1-\frac{V^{(p)}}{V}]\frac{V}{V^{(p)}}
  \label{eq:appD:NWHELP}
\end{equation}
The local number density of water molecules in solution, $n_w(\bm{r},t)$, is
defined like in Eq.~(S16):
\begin{equation}
   n_{w}(\bm{r},t)=n_{0}-\int n_{sol}(\bm{r'},t)n_{w}^{(p)}(\bm{r}-\bm{r'},t)\, \mathrm{d}\bm{r'}
  \label{eq:appD:nwater}
\end{equation}
where $n_w^{(p)}(\bm{r},t)$ is equal to $n_0$ if $\bm{r}$ lies within the excluded volume
and zero otherwise. According to Eq.~(\ref{eq:appD:nwater}), $n_w(\bm{r},t)$ is zero inside the excluded
volume. The water-water pair correlation function, $G_w(\bm{r},t)$,
is:
\begin{equation}
  G_{w}(\bm{r},t)=\frac{1}{N_{w}}\int n_{w}(\bm{r'})n_{w}(\bm{r'}+\bm{r},t)\, \mathrm{d}\bm{r'}
  \label{eq:appD:gwater}
\end{equation}
The substitution of Eq.~(\ref{eq:appD:nwater}) into Eq.~(\ref{eq:appD:gwater}) yields:
\begin{eqnarray}
G_{w}(\bm{r},t)=n_{0}^{2}/n_{w}+(2n_{0}(1-\frac{n_{0}}{n_{w}})) & + &  \frac{MN_{sol}}{N_{w}} \int
 G_{sol}^{self(p)}(\bm{u},t)G_{sol}^{self}(\bm{r}-\bm{u},t)  \, \mathrm{d}\bm{u}   \nonumber \\ & +  & 
  {\int  G_{sol}^{dist(p)}(\bm{u},t)G_{sol}^{dist}(\bm{r}-\bm{u},t) \, \mathrm{d}\bm{u} }
  \label{eq:appD:gwaterexpand}
\end{eqnarray}
The functions $G_{sol}^{self (p)}(\bm{u}, t)$ and $G_{sol}^{dist (p)}(\bm{u}, t)$ are defined as
in Eqs.~(\ref{eq:appD:grtselfinternal}), (\ref{eq:appD:grtdistinternal}), but they describe the time-dependent correlations
between the infinitesimal volume elements of the excluded volumes (i.e., 
between the CMs of water molecules, if the excluded volumes were filled
with water). It follows (see Eq.~(\ref{eq:appD:NWHELP})):
\begin{eqnarray}
G_{w}(\bm{r},t)=  \frac{n_{0}^{2}}{n_{w}} \left ( 1-\frac{2V^{(p)}}{V}+\frac{ \nu^{(p)} n_{sol}}{n_{0}}\int
G_{sol}^{self(p)}(\bm{u},t)G_{sol}^{self}(\bm{r}-\bm{u},t) \, \mathrm{d}\bm{u}  \right.   \nonumber \\
 +  \left.    \frac{ \nu^{(p)} n_{sol}}{n_{0}}\int G_{sol}^{dist(p)}(\bm{u},t)G_{sol}^{dist}(\bm{r}-\bm{u},t)\, \mathrm{d}\bm{u} \right ) 
  \label{eq:appD:gwaterexpand2}
\end{eqnarray}

The solute-water time-dependent pair correlation function, $G_{sol-w}(\bm{r}, t)$, is:
\begin{equation}
G_{sol-w}(\bm{r},t)=\frac{1}{N_{sol}}\int
n_{sol}(\bm{r'})n_{w}(\bm{r}+\bm{r'},t)\, \mathrm{d}\bm{r'}  
  \label{eq:appD:gsolw}
\end{equation}
where $n_{sol}(\bm{r}, t)$ is the local time-dependent number density of the
CM's of solute molecules. Putting Eq.~(\ref{eq:appD:nwater}) into 
 Eq.~(\ref{eq:appD:gsolw}) and substituting $\bm{u} = \bm{r} + \bm{r'} - \bm{r''}$, yields:
\begin{equation}
  G_{sol-w}(\bm{r},t)=n_{0}-\frac{1}{N_{sol}}\int
\, \mathrm{d}\bm{u}\langle n_{w}^{(p)}(\bm{u},t)\rangle _{\Omega }\int \, \mathrm{d}\bm{r''}\langle
n_{sol}(\bm{r''},t)n_{sol}(\bm{u}-\bm{r}+\bm{r''})\rangle
  \label{eq:appD:gsolwexpand}
\end{equation}
It follows from Eq.~(\ref{eq:appD:grtdefinition}):
\begin{equation}
G_{sol-w}(\bm{r},t)=n_{0}-\int \langle n_{w}^{(p)}(\bm{u},t)\rangle _{\Omega
}G_{sol}(\bm{r}-\bm{u},t)\, \mathrm{d}\bm{u}
  \label{eq:appD:gsolwexpand2}
\end{equation}
%
%  INTERMEDIATE SCATTERING FUNCTIONS
%
\subsection{\label{subsec:uniformiqt}Water-water and solute-water intermediate scattering functions in the UFA} 
%
%  INTERMEDIATE SCATTERING FUNCTIONS
%
%%%
Applying the convolution theorem of Fourier transformation to $G_w(\bm{r},t)$ given by Eq.~(\ref{eq:appD:gwaterexpand2})  and 
to $G_{sol-w}(\bm{r},t)$ given by Eq.~(\ref{eq:appD:gsolwexpand2}), after 
 averaging over $\bm{Q}$-orientaions, 
 we get the corresponding intermediate scattering functions, $I^{coh}_{tr~w}(Q,t)$ and $I_{tr~sol-w}(Q,t)$, respectively.
 Specifically, $I^{coh}_{tr~w}(Q,t)$ is (omitting the term containing $\delta (Q)$):
\begin{equation}
  I^{coh}_{tr~w}(Q,t)=\frac{n_{sol}}{n_{w}} ( A^{0(p)}(Q)I_{tr~sol}^{dist}(Q,t)+\xi
^{(p)}(Q,t)I_{tr~sol}^{self}(Q,t)  )
  \label{eq:appD:itrw}
\end{equation}
Apart from the factor $n_{sol}/n_w$, Eq.~(\ref{eq:appD:itrw}) is the same as Eq.~(6) in
Sears's paper on the scattering by molecules in
liquids \cite{SEARS1967P3} (except that coherent scattering lengths and cross-sections
will not appear in the formulae for $A^{l(p)}(Q)$). Indeed, the
time-dependent spatial correlations between the nuclei in a reorienting
polyatomic molecule are analogous to the correlations between the
infinitesimal volume elements in the volume excluded by a reorienting
solute molecule. The model of continuous rotational diffusion yields $\xi ^{(p)}(Q,t)$ \cite{SEARS1967P3}:
\begin{equation}
\xi ^{(p)}(Q,t)=\sum _{l=0}^{\infty }(2l+1)A^{l(p)}(Q)e^{-l(l+1)D_{r~sol}t}
  \label{eq:appD:xirotation}
\end{equation}
and the coefficients $A^{l(p)}(Q)$ are (see Eq.~(\ref{eq:appA:AlQcoef}) for notations):
\begin{equation}
A^{l(p)}(Q)=\int \int
j_{l}(Qr_{1})j_{l}(Qr_{2})P_{l}(\cos \theta
_{12})n_{w}^{(p)}(\bm{r}_{1})n_{w}^{(p)}(\bm{r}_{2}) \, \mathrm{d}\bm{r}_{1}  \mathrm{d}  \bm{r}_{2}
  \label{eq:appD:alpcoefficients}
\end{equation}
where $n^{(p)}_w(\bm{r})$ is equal to the pure water number density ($n_0$) if $\bm{r}$ lies inside the
excluded volume and zero otherwise.

$I_{tr~sol-w}(Q,t)$ is (omitting the term containing $\delta (Q)$):
\begin{equation}
  I_{tr~sol-w}(Q,t)=-N^{(p)}(Q)I^{coh}_{tr~sol}(Q,t)
  \label{eq:appD:itrsolw}
\end{equation}
where $N^{(p)}(Q)$ is:
\begin{equation}
N^{(p)}(Q)=\int {\frac{\sin
(Qr)}{Qr}}n_{w}^{(p)}(\bm{r}) \, \mathrm{d} \bm{r}
  \label{eq:appD:NPQ}
\end{equation}

From Eq.~(\ref{eq:appD:itrw}) and Eq.~(\ref{eq:appD:itrsolw})  the scattering functions given by  Eqs.~(\ref{eq:sqwwater_uniform1})-(\ref{eq:sqwwater_uniform3}) 
 and Eq.~(\ref{eq:sqwcross_uniform}), respectively,  
are obtained by Fourier transformation.

%
%
%
% BIBLIOGRAPHY
%
%
  % Create the reference section using BibTeX:
%\bibliography{mybib}
%Merlin.mbs v4.21 2009-07-09.
%

%
%
\end{document}